%% file: main.tex
\documentclass[acmsmall]{acmart}

\input{packages_macros}

\setcopyright{rightsretained}
\acmDOI{10.1145/3643755}
\acmYear{2024}
\copyrightyear{2024}
\acmSubmissionID{fse24main-p517-p}
\acmJournal{PACMSE}
\acmVolume{1}
\acmNumber{FSE}
\acmArticle{29}
\acmMonth{7}
\received{2023-09-29}
\received[accepted]{2024-01-23}

\begin{document}


\title{Unprecedented Code Change Automation: The Fusion of LLMs and Transformation by Example}

 




\author{Malinda Dilhara}
\orcid{0000-0001-5471-7179}
\affiliation{%
  \institution{University of Colorado}
  \city{Boulder}
  \country{USA}
}
\email{malinda.malwala@colorado.edu}

\author{Abhiram Bellur}
\orcid{0009-0008-7048-4406}
\affiliation{%
  \institution{University of Colorado}
  \city{Boulder}
  \country{USA}
}
\email{Abhiram.Bellur@colorado.edu}

\author{Timofey Bryksin}
\orcid{0000-0001-9022-3563}
\affiliation{%
  \institution{JetBrains Research}
  \city{Limassol}
  \country{Cyprus}
}
\email{timofey.bryksin@jetbrains.com}

\author{Danny Dig}
\orcid{0000-0001-5046-2017}
\affiliation{%
  \institution{JetBrains Research, University of Colorado}
  \city{Boulder}
  \country{USA}
}
\email{danny.dig@colorado.edu}

\begin{abstract}
Software developers often repeat the same code changes within a project or across different projects. 
These repetitive changes are known as ``code change patterns'' (\cpats).
Automating \cpats is crucial to expedite the software development process.
While current Transformation by Example (TBE) techniques can automate \cpats, they are limited by the quality and quantity of the provided input examples. 
Thus, they miss transforming code variations that do not have the exact syntax, data-, or control-flow of the provided input examples, despite being semantically similar.
Large Language Models (LLMs), pre-trained on extensive source code datasets, offer a potential solution. 
Harnessing the capability of LLMs to generate semantically equivalent, yet previously unseen variants of the original \cpat could significantly increase the effectiveness of TBE systems.


In this paper, we first discover best practices for harnessing LLMs to generate code variants that meet three criteria: correctness (semantic equivalence to the original \cpat), usefulness (reflecting what developers typically write), and applicability (aligning with the primary intent of the original \cpat). 
We then implement these practices in our tool \textsc{PyCraft}, which synergistically combines static code analysis, dynamic analysis, and LLM capabilities. 
By employing chain-of-thought reasoning, \textsc{PyCraft} generates variations of input examples \emph{and} comprehensive test cases that identify correct variations with an F-measure of 96.6\%. 
Our algorithm uses \fIteration to expand the original input examples by an average factor of 58x.
Using these richly generated examples, we inferred transformation rules and then automated these changes, resulting in an increase of up to \maxOpportunitiesRatioOverBaseLine, with an average increase of \opportunitiesRatioOverBaseLine in target codes compared to a previous state-of-the-art tool that relies solely on static analysis.
We submitted patches generated by \textsc{PyCraft} to a range of projects, notably esteemed ones like \textit{microsoft/DeepSpeed} and \textit{IBM/inFairness}. 
Their developers accepted and merged \acceptedRatioOfCPATS the \pullCPATInstances CPAT instances submitted through \numberOfPullReq pull requests.
This confirms the usefulness of these changes.
\end{abstract}



\maketitle

\section{Introduction}
Software developers frequently change code to improve performance, manage resources efficiently, integrate  new libraries, etc.
Throughout this process, they frequently repeat  identical or similar code modifications~\cite{hindle2016onthenaturalness,negara2014codechanges,barr2014plasticsurgary,nguyen2019cpatminer,RobertMiningBillions2014}. 
These repetitions stem from the adoption of shared coding idioms~\cite{Phanudom2020Teddy,MiltiadisMiningIdiomFSE2014s,coplien1991advanced,Zhang2022IdiomaticPython}, adherence to common best practices~\cite{rcaptminer2022ICSE,JavaStreams}, and the need to tackle similar programming challenges~\cite{inferrule,serrano2020spinfer}. 
Such repeated changes occur at a fine-grained level, frequently appearing within specific methods, and consistently retaining the same semantics.

For example, \Cref{fortonumpy} shows such a repeated change in project \textit{NifTK/NiftyNet}, an open-source convolutional neural network platform. 
The developers replaced a \smcode{for} loop that summed \smcode{elements} of a list with the more efficient \smcode{numpy.sum} function, which is a best practice and improves performance.
The performance gain is attributed to several factors, including NumPy's C implementation, vectorization, memory efficiency, optimized algorithms, and parallel processing support.
This change involves specific programming idioms and is localized to a particular method within the project. 
This change recurs at multiple locations across various commits, prompting previous researchers~\cite{pyevolve2022ICSE,nguyen2019cpatminer,rcaptminer2022ICSE} to identify these recurrent changes as \textit{code change \textbf{patterns}} (\cpats). 

\begin{figure}[b]
\vspace{-5mm}
\begin{minipage}{0.45\textwidth}
\begin{lstlisting}[language=Python, caption=Commit c8b28432 in GitHub project NifTK/NiftyNet: Replace \smcode{for} loop with \smcode{numpy.sum}
\label{fortonumpy}]
-*\colorbox{tearose}{result = 0}*
-*\colorbox{tearose}{for elem in elements:}*
-   *\colorbox{tearose}{result = elem + result}*
+*\colorbox{teagreen}{result = numpy.sum(elements)}*
\end{lstlisting}
\end{minipage}%
\hspace{5mm}
\begin{minipage}{0.45\textwidth}
\begin{lstlisting}[style=mystyle,language=Python, caption=GitHub repository CDE-GAN/models employs a \smcode{for} loop for calculating the sum of an array similar to the tranformed code in \Cref{fortonumpy}\label{lst:npsumtype4}]
loss = 0    
for i in range(len(losses)):
    loss += losses[i]    
\end{lstlisting}
\end{minipage}
\vspace{-5mm}
\end{figure}


Developers often perform these changes manually~\cite{nguyen2019cpatminer,rcaptminer2022ICSE}: they first must identify potential target code sites and then  apply the required syntax transformations. 
This manual process is time-consuming, tedious, and error-prone due to two main reasons: 
\begin{enumerate*}[label=(\roman*)]
\item identifying all potential target sites to apply \cpats is challenging, as they may be deeply embedded within the code, exhibiting syntactic variations along with variations in data- and control-flow, and
\item ensuring consistency and correctness in applying identical changes across multiple locations in the codebase is challenging, as developers must reason about syntactic variations.
\end{enumerate*}
To address these challenges and enhance developer productivity, researchers have employed \PBE (TBE)~\cite{MLCatchUp,Stefanus2021Stefanus,Haryono2022AndroEvolve,Hora2013MiningRules,Bader2019Getafix,inferrule,MengSystematicEditingPLDI2011,MiltnerOnTheFly2019,Rolim:Revisar,DagenaisRecommending2011,AppEvolveFazziniAST2019,HenkelCatchUp2005,Meditor:ICPC:2019,A3LamotheTSE2022,A3LamotheTSE2022,Yu2017ASE,BlinkFill} techniques to automate code changes.
These techniques infer transformation rules from before-and-after versions of code changes and use these inferred rules to automatically transform new target code sites that exhibit syntactical similarities and have a program structure similar to the original code change.
These approaches are effective in API migrations, e.g., replacing obsolete API calls with modern ones from the Android SDK~\cite{APIMigrator,AppEvolveFazziniAST2019,Meditor:ICPC:2019,Xiang2021APIFix,Stefanus2021Stefanus,A3LamotheTSE2022}, type migrations in Java~\cite{inferrule}, and API migrations in Linux systems~\cite{serrano2020spinfer}.

{Despite their successes, existing TBE  techniques face challenges when handling more complex coding idioms, where there can be numerous potential target sites that are semantically equivalent, yet differ in terms of syntax and data- and control-flow. 
These techniques are limited to transforming target codes that \emph{perfectly resemble} the input example code change and struggle with variations beyond those specific examples. We call these \emph{previously unseen variants}. 
Unseen variants can be of two types:
\begin{enumerate*}[label=(\roman*)]
\item syntax variants, and/or
\item data- or control-flow variants.
\end{enumerate*}
\Cref{lst:npsumtype4} shows an example of syntax variant where the code computes the sum of elements in the list \smcode{losses} by using a 
different syntax involving \smcode{len} and \smcode{range}. It is also a data-flow variant  through  list indexing and accumulation in the variable \smcode{loss}. 
Notably, this target code would not be identified or transformed by existing techniques that inferred  transformation rules based on \Cref{fortonumpy} as input. 
Our initial analysis uncovered 50 other ways to compute the sum of elements, all of which are unseen variants of \Cref{fortonumpy}. They could all benefit from and be sped up by  transforming them to \smcode{np.sum}.
Sadly, existing techniques fail to identify and transform these unseen variants.}



Recent advancements,  \textit{PyEvolve}~\cite{pyevolve2022ICSE} and \textit{Spinfer}\cite{serrano2020spinfer}, are effective in handling certain unseen data- or control-flow variants. 
However, none of the existing tools are equipped to handle syntax variants.
While \textit{Spinfer}~\cite{serrano2020spinfer} addresses control variations through the use of the "..." operator to represent arbitrary statements between specific statements in input examples, it relies on multiple input examples to learn the potential locations for inserting "..." in the rule. 
This reliance on examples can be problematic if not all potential locations for arbitrary statements are covered, limiting its effectiveness in handling many unseen variants. On the other hand, \textit{PyEvolve}~\cite{pyevolve2022ICSE} excels in automating unseen variants by employing a graph-based approach to support variations not exemplified in the input examples, specifically those concerning data- and control-flow. However, this approach is limited to automating simple data and control variations, such as reassigning values to other variables, and it cannot handle more complex variations, as seen in \Cref{lst:npsumtype4}, where list elements are accessed using indexing (\smcode{loss += losses [i]} in \Cref{lst:npsumtype4}). 

To advance the frontier for \PBE systems, we designed, implemented, and evaluated a novel approach and a tool,   \tool. 
It successfully automates unseen variants, even those with \emph{completely different syntax}.
\tool harnesses the creativity of Large Language Models (LLMs) to generate many syntactical variations for a given code idiom.
LLMs are robust machine learning models that are trained on vast datasets, which encompass source codes as well as documents related to software development.
Models like GPT-3.5~\cite{brown2020language} and GPT-4~\cite{rae2023gpt} generate coherent and contextually relevant code snippets in response to given prompts. 
Researchers demonstrated LLMs' versatility across various software engineering tasks, including code completion~\cite{Matteo2022TSE}, 
refactoring~\cite{pomian2024together,pomian2024assist}, code summarization~\cite{feng-etal-2020-codebert,CITE_AleksandraASE23}, and bugs reproduction~\cite{Feng2024ICSE}.

In \tool, we employ few-shot learning to generate  unseen variants for the before-part of a given input \cpat. 
Our initial analysis found that \LLMerrorrate of the LLM-generated variants are either erroneous or not semantically equivalent to the original \cpat and cannot be used. 
Therefore,
we discovered three criteria that the generated variants must meet: (i) semantically similar to the original input (i.e., correct), (ii) 
practicality in terms of what developers typically write (i.e., useful), and (iii) aligned with the structural intent of the original \cpat (i.e., applicable). 
To achieve this, we carefully fine-tune the hyper-parameters of the LLM, and conduct comprehensive automatic checks, including static code validation, to ensure correctness and verify that the variants adhere to the original \cpat's structural intent.
We also perform dynamic analysis through automatically generated test cases to ensure conformity with the desired behavior.
To optimize LLM performance, we fine-tune the hyper-parameters by discovering best practices for generating variants and test cases, finding that higher temperature values are effective for dynamic analysis-focused test case generation, while intermediate temperatures help reduce non-useful variants.
This thorough process guarantees the reliability and effectiveness of \tool.
Finally, \tool infers transformation rules that are used to apply \cpat to new target codes including unseen variants, significantly enhancing state-of-the-art TBE techniques and further advancing their automation capabilities.


We conducted a comprehensive evaluation of \tool to assess its effectiveness in generating variants and the usefulness of the resulting code changes.
We utilized \cpats mined from open-source repositories as input for \tool and observed that it could generate raw variations of up to \LLMrawVariants per \cpat, with an average \CPATtoOriginalVariation per \cpat.
To obtain high-quality applicable variants, we use a combination of static and dynamic analysis to validate the raw variants. 
\tool consistently generated applicable variants at an average \CPATtoVariatoin per \cpat while effectively eliminating inapplicable variants, on average \CPATtoRemovedVariatoin per \cpat.
This not only demonstrates its proficiency in identifying irrelevant variants but also its ability to successfully infer previously unattainable rules using state-of-the-art techniques.
Furthermore, to quantitatively compare \tool and the previous state-of-the-art, we conducted a comparison analysis using \tool and \textit{PyEvolve}, the leading tool for automating unseen variations.
\tool exceeded the baseline by enabling an average of \opportunitiesRatioOverBaseLine additional code transformation instances per \cpat; 
these would have been missed by prior tools. 
Furthermore, to assess the usefulness of the generated variations, we submitted pull requests to highly-rated projects including \textit{microsoft/DeepSpeed} and \textit{IBM/inFairness}, totaling \pullCPATInstances CPAT instances. 
At the time of this writing, developers have accepted \acceptedInstances ~(\acceptedRatioOfCPATS) of the CPAT instances, submitted through \numberOfPullReq pull requests.
This confirms the practical value of the transformations performed by \tool for the real-world developers.

This paper makes the following key contributions:
\begin{enumerate}[label=\bfseries (\arabic*),wide, labelwidth=!, labelindent=0pt]
\item We pioneer a new approach that utilizes LLMs in TBE to generate unseen variants, thereby automating code changes that were once unattainable with existing TBE techniques.
\item We provide best practices for using LLMs to generate code variations and their test cases.
\item We designed, implemented, and evaluated these ideas in a new tool, \tool. We perform comprehensive experiments, including a performance evaluation of \tool and a detailed qualitative analysis, to demonstrate the capabilities of LLMs in generating variations and the effectiveness of our technique for selecting the applicable ones for automation. Moreover, we conduct a user study to further validate the usefulness of our approach.
\item Our tool and evaluation dataset is open-source and available for others to reuse~\cite{artifacts}.
\end{enumerate}

\input{MotivatingExample}

\input{Technique}

\input{Evaluation}

\input{ThreatsToValidity}

\input{RelatedWork}

\section{Conclusions and future work}
{To accelerate the software development process, it is essential to automate code changes (\cpats). 
Given the wide variety of \cpats, we can not encode the analysis and transformation for each \cpat, and instead we turn to using TBE systems.
So far, progress on using TBE systems to automatically apply \cpats has been stifled by the requirement that new target codes exhibit syntax, data-, and control-flow similar to the original \cpat. 
Our novel approach and tool, \textit{\tool}, pioneers the synergistic integration of LLMs with TBE systems, and leverages the LLMs' strengths to provide breakthroughs for previous limitations. 

Prior to our recent advancements, TBE systems would mine \cpats from aging code bases, only to fall prey to software evolution that would render these \cpats obsolete and less applicable over time. 
During our extensive evaluation, we observed that \textit{\tool} successfully transformed even target codes that employ the latest language constructs, coding idioms, and library versions. To do this, it generates code variants that employ constructs and features not seen during \cpat mining.
Achieving this level of compatibility was unfeasible with earlier tools. 
By staying \emph{perpetually young} and up to date, \textit{\tool} transcends software evolution and can revolutionize code change automation.  

We anticipate significant usage of LLMs for software engineering (SE) tasks in the coming years. Using \tool, we demonstrated this potential in code change automation, achieving an increase in code transformations of \opportunitiesRatioOverBaseLine over the baseline. Moreover, our research confirmed LLMs' creativity \emph{and} their propensity to hallucinate. This highlights the need for targeted research and  development of validation techniques to make our SE tools immune to LLM hallucinations. 
We hope that the techniques we developed in this research inspire others to build upon and move the field forward.

\section*{Acknowledgements}
{We thank the ML Methods in Software Engineering Lab at JetBrains Research, and the FSE-2024 reviewers for their insightful and constructive feedback for improving the paper.
This research was partially funded through the NSF grants CNS-1941898, CNS-2213763, and the Industry-University Cooperative Research Center on Pervasive Personalized Intelligence.}


\bibliographystyle{ACM-Reference-Format}
\bibliography{references}

\end{document}

%% file: packages_macros.tex
\usepackage{amsmath,amsfonts}
\usepackage{algorithmic}
\usepackage{algorithm}
\usepackage{array}
\usepackage{textcomp}
\usepackage{stfloats}
\usepackage{url}
\usepackage{verbatim}
\usepackage{graphicx}
\usepackage{subfigure}
\usepackage{xspace}
\usepackage{hyperref}
\usepackage{listings}
\usepackage{makecell}
\usepackage{booktabs}
\usepackage{algorithm}
\usepackage{threeparttable}
\usepackage{longtable}
\usepackage{tabularray}
\usepackage{cleveref}
\hypersetup{hidelinks}
\usepackage{color, colortbl}
\usepackage{balance}
\usepackage[inline]{enumitem}
\usepackage{calligra}
\usepackage{amsmath,amsthm}
\usepackage{tikz}
\usetikzlibrary{shapes.geometric, arrows}
\usetikzlibrary{positioning}
\usepackage{pifont}
\usepackage{tcolorbox}
\usepackage{makecell}
\usepackage{fancybox}

\makeatletter

\lstdefinestyle{mystyle}{
    backgroundcolor=\color{backcolour},   
    basicstyle=\footnotesize,
    commentstyle=\color{codegreen},
    keywordstyle=\color{magenta},
    numberstyle=\tiny\color{codegray},
    stringstyle=\color{codepurple},
    basicstyle=\ttfamily\footnotesize,
    breakatwhitespace=true,         
    breaklines=true,                 
    captionpos=t,                    
    keepspaces=true,                 
    numbers=left,                    
    numbersep=5pt,                  
    showspaces=false,                
    showstringspaces=false,
    escapeinside={*}{*},
    showtabs=false,                  
    tabsize=2,
    frame=none,
    postbreak=\mbox{\textcolor{red}{$\hookrightarrow$}\space},
}
\lstdefinestyle{tablecode}{
  basicstyle=\fontsize{5}{6}\selectfont,
  breaklines=true,
  backgroundcolor=\color{white},
  stringstyle=\color{codepurple},
  numbers=none,
  frame=none,
  lineskip=-4pt,
  xleftmargin=0pt,
  framexleftmargin=0pt,
  framexrightmargin=0pt,
  framexbottommargin=0pt,
  framextopmargin=0pt,
  showspaces=false,  
  showtabs=false,   
  showstringspaces=false, 
  escapeinside={|}{|},
  mathescape=true,
  morekeywords={value},
  deletekeywords=int,
}

\definecolor{tearose}{rgb}{0.96, 0.76, 0.76}
\definecolor{ruddypink}{rgb}{0.88, 0.56, 0.59}
\definecolor{pastelred}{rgb}{1.0, 0.41, 0.38}
\definecolor{javapurple}{rgb}{0.5,0,0.35}
\definecolor{teagreen}{rgb}{0.82, 0.94, 0.75}
\definecolor{codegreen}{rgb}{0.1,0.9,0}
\definecolor{yellow-green}{rgb}{0.6, 0.8, 0.2}
\definecolor{codegray}{rgb}{0.5,0.5,0.5}
\definecolor{codepurple}{rgb}{0.58,0,0.82}
\definecolor{backcolour}{rgb}{0.95,0.95,0.92}
\definecolor{blueannoback}{RGB}{234,242,250}
\definecolor{myblue}{rgb}{0, 0, 0.75}
\definecolor{lightgray}{gray}{0.85}
\definecolor{Blue}{rgb}{0.1, 0.1, 0.2}

\newcommand{\code}[1]{\texttt{\fontsize{9.5}{10}\selectfont #1}}
\newcommand{\smcode}[1]{\texttt{\fontsize{7.5}{8}\selectfont #1}}
\newcommand{\cpat}{\textit{CPAT}\xspace}
\newcommand{\cpats}{\textit{CPATs}\xspace}

\newcommand{\PBE}{``Transformation by Example''\xspace}

\theoremstyle{definition}
\newtheorem{definition}{Definition}[section]

\newcommand{\tool}{\textit{{PyCraft}}\xspace}
\newcommand{\toolit}{\textit{PyCraft}\xspace}
\newcommand{\templateVar}[1]{\textcolor{blue}{\smcode{#1}}}

\newcommand{\VTone}{\textit{VT1}\xspace}
\newcommand{\VTtwo}{\textit{VT2}\xspace}
\newcommand*\circled[1]{\tikz[baseline=(char.base)]{
            \node[shape=circle,draw,inner sep=0.8pt] (char) {#1};}}

\newcommand{\rqfourVI}{5284\xspace}     
\newcommand{\rqfourVC}{2780\xspace}

\newcommand{\resultbox}[1]{
\begin{center}
\begin{tcolorbox}[colback=gray!20!white,colframe=gray!60!black,left=0pt,right=0pt,top=0pt,bottom=0pt]
    \begin{minipage}{.99\textwidth}
    #1
    \end{minipage}
\end{tcolorbox}
\end{center}
}

\newcommand{ \LLMerrorrate}{65.5\%\xspace}

\newcommand{ \opportunitiesRatioOverBaseLine}{14x\xspace}  
\newcommand{ \maxOpportunitiesRatioOverBaseLine}{39x\xspace} 
\newcommand{ \minOpportunitiesRatioOverBaseLine}{4x\xspace}

\newcommand{\CPATtoOriginalVariation}{\textit{459}\xspace}  
\newcommand{\CPATtoRemovedVariatoin}{\textit{748}\xspace}  
\newcommand{\CPATtoVariatoin}{\textit{58}\xspace}  
  
\newcommand{\LLMrawVariants}{584\xspace}

\newcommand{\rqSixProjects}{40\xspace}  
\newcommand{\pullCPATInstances}{86\xspace}  
\newcommand{\pullUpdatedFiles }{124\xspace}  
\newcommand{\pullUpdatedLOC }{590\xspace}   
\newcommand{\numberOfPullReq }{44\xspace}

\newcommand{\acceptedRatioOfCPATS}{83\%\xspace}  
\newcommand{\acceptedInstances}{72\xspace}  
\newcommand{ \rejectedPulls}{4\xspace}

\newcommand{\trendone}{\textit{transform to context managers}\xspace}
\newcommand{\trendtwo}{\textit{dissolve \smcode{for} loops to domain-specific abstractions}\xspace}
\newcommand{\trendthree}{\textit{update API usage}\xspace}
\newcommand{\trendfour}{\textit{use advanced language features}\xspace}

\newcommand{\fIteration}{feedback iteration\xspace} 
\newcommand{\fIterations}{feedback iterations\xspace} 
\newcommand{\FIteration}{Feedback Iteration\xspace}

%% file: MotivatingExample.tex
\section{Motivating Examples\label{sec:moticationexample}}
To demonstrate the challenges faced by current techniques that use \PBE on \cpats, we employ real-world code changes. 
First we show a scenario when existing techniques exhibit a high recall rate, effectively automating such changes. 
Then, we present two complex scenarios when existing techniques fail to automate code changes, whereas \toolit is successful.

\begin{figure}
\begin{minipage}{0.53\textwidth}
\begin{lstlisting}[language=Python, caption=Tranformation rule for the code change in \Cref{fortonumpy}\label{lst:ruleNumpySum},style=tablecode]
:[[v0]] = 0                          $\Rightarrow{}$   :[[v0]]=numpy.|sum|(:[[v2]])
for :[[v1]] in :[[v2]]: 
    :[[v0]]  = :[[v0]] + :[[v1]]
\end{lstlisting}
\end{minipage}%
\hspace{3mm}
\begin{minipage}{0.44\textwidth}
\begin{lstlisting}[style=mystyle,language=Python, caption=The repository Yolov uses a \smcode{for} loop to compute the sum of an array\label{control_sum} ]
temp_list = [0] + int_list
for i in range(1, len(temp_list)):
    temp_list[i] += temp_list[i - 1]
count = temp_list[-1]
\end{lstlisting}
\end{minipage}
\vspace{-5mm}
\end{figure}

\Cref{fortonumpy} shows an example of a \cpat mined from  the project \textit{NifTK/NiftyNet}.
The developer transforms the \smcode{for} loop that computes the sum to \smcode{np.sum()}, a highly optimized API from the library \textit{NumPy}.
The transformation is represented by the rewrite rule shown in Listing \ref{lst:ruleNumpySum}.
The rule, following ComBy syntax~\cite{comby}, has a left hand side (LHS -- indicating the ``before'' change) and a right side (indicating the ``after''  change) separated by an arrow.
Both sides of the rule contain Python statements with template variables (e.g., \templateVar{:[[v0]]}) that bind to AST nodes in the actual source code (e.g., \templateVar{:[[v0]]} binds to \smcode{result}).
The right side represents the code fragments that replace the left side.
These rules can be applied to transform any target code that shares the same AST structure as the code presented in \Cref{fortonumpy}, irrespective of any variations in variable names.
Notably, existing TBE techniques infer this rule and automate the transformation of structurally similar target codes.

However, many real-world \cpats involve semantically equivalent different variations~\cite{pyevolve2022ICSE,rcaptminer2022ICSE}.
These variations can be identified in two different ways.
The first type, Variant Type 1 (\VTone), consists of code fragments that possess different syntax but exhibit similar semantics, resembling Type 4 clones~\cite{HiteshSourcererCC2016}.
The second type, Variant Type 2 (\VTtwo), comprises code fragments that are syntactically equivalent but differ in terms of data- and control-flow.


\Cref{lst:npsumtype4} is an example of \VTone. It iterates over a list to compute the sum of elements in the array \smcode{int\_list} similar to \Cref{fortonumpy}. 
However, unlike \Cref{fortonumpy},  it utilizes the \smcode{len()} function to determine the length of the list \smcode{int\_list}. This length value is then used with the \smcode{range()} function to generate a sequence of numbers. 
The loop iterates over this sequence, accessing each indexed element of \smcode{int\_list} for performing the sum operation. 
This code fragment comes from project CDE-GAN/models and is semantically similar to the \Cref{fortonumpy}. 
However, it differs syntactically, making it a Type-4 clone of the \Cref{fortonumpy}. 
Notably, this target code should be similarly transformed into the \smcode{np.sum()} function, as demonstrated in the \cpat given in \Cref{fortonumpy}.
The rule shown in \Cref{lst:newrule} is essential for transforming the code presented in \Cref{lst:npsumtype4}.
Since \Cref{lst:npsumtype4} was not seen as a change exemplar, existing TBE techniques are unable to infer the rule in \Cref{lst:newrule}, resulting in their failure to transform this new target site. In contrast, \toolit succeeds in this task.

\Cref{control_sum} represents a \VTtwo variant that computes the sum of elements in the \smcode{int\_list} array, similar to \Cref{fortonumpy}.
It differs from \Cref{fortonumpy} by assigning the initial list to the temporary list \smcode{temp\_list} and computing the cumulative sum of the array elements, rather than using an accumulator variable. 
The last element in the list is the sum of all the elements.
This variation in data-flow classifies it as a data-flow variant, thereby denoting it as a \VTtwo variant.
Among the existing TBE tools, \textit{PyEvolve}~\cite{pyevolve2022ICSE} can partially automate these \VTtwo variants. 
It captures control-flow variations that include unrelated nodes among the pattern's statements, as well as data-flow variations such as reassigning a variable to another variable that is not exemplified in the \cpat. 
However, it falls short in automating the transformation of unseen variations that contain complex data-flow relations, such as those observed in \Cref{control_sum}: 
\begin{enumerate*}[label=(\roman*)]
\item reassigning the cumulative sum to the array elements, and
\item differentiating between the use of the ``addition assignment'' operator (\smcode{temp\_list[i] += temp\_list[i-1]}) and the ``addition and assignment'' operators (\smcode{result = elem + result}).
\end{enumerate*}
Therefore, existing techniques can not fully automate complex \VTtwo variations; whereas \tool succeeds.

\begin{figure}
\begin{minipage}{0.41\textwidth}
\begin{lstlisting}[style=mystyle,language=Python, caption=An unseen variation generated by LLM for the LHS in \Cref{fortonumpy}\label{lst:nonuseful} ]
result = 0
for i in sorted(elements):
    result += i
\end{lstlisting}
\end{minipage}%
\hspace{3mm}
\begin{minipage}{0.55\textwidth}
\begin{lstlisting}[language=Python, caption=Transformation rule that need to be inferred to transform code in \Cref{lst:npsumtype4}\label{lst:newrule},style=tablecode]
:[[v0]] = 0
for :[[v1]] in range(len(:[[v2]])):     $\Rightarrow{}$  :[[v0]]=numpy.|sum|(:[[v2]])
    :[[v0]]  += :[[v2]][:[[v1]]] 
\end{lstlisting}
\end{minipage}
\vspace{-6mm}
\end{figure}

The key idea behind \tool is to generate \VTone and \VTtwo unseen code variations using an LLM.  
However, not all of the generated variants can be directly used for rule inference.
Some may contain syntax errors, import issues, incorrect types, or lack semantic equivalence. 
Our empirical evaluations show that these erroneous variants can account for as much as 76\% of the total generated variants. 
Even if 24\% of them are correct, not all of these correct variants are suitable for the final rule inference, as certain semantically equivalent correct variants may not be useful.
For example, the variant in \Cref{lst:nonuseful} performs an unnecessary sorting operation that an actual developer would not typically perform, making it a \emph{not-useful} variant. 
Our empirical evaluations indicate that these not-useful variants can constitute up to 80\% of the total correct variants and significantly impact the tool's performance. Therefore, it is essential to reduce them while increasing the useful variants.

Even when certain variants are indeed useful, some may still deviate from the original structural intent. 
For example, the variant \smcode{sum(elements)}, generated by LLM to calculate the sum of elements in an array, is semantically equivalent to \Cref{fortonumpy} and useful. 
However, it does not iterate a collection as in \Cref{fortonumpy}, deviating from the intended structural intent, making it not applicable. 
We observed that these non-applicable variants can account for up to 70\% of the useful variants. 
Thus, \tool selects correct, useful, and applicable variants (e.g., 
\Cref{lst:npsumtype4}) for the final rule inference.

%% file: Technique.tex
\section{technique}
\begin{figure}
\includegraphics[width=0.9\textwidth,keepaspectratio=true]{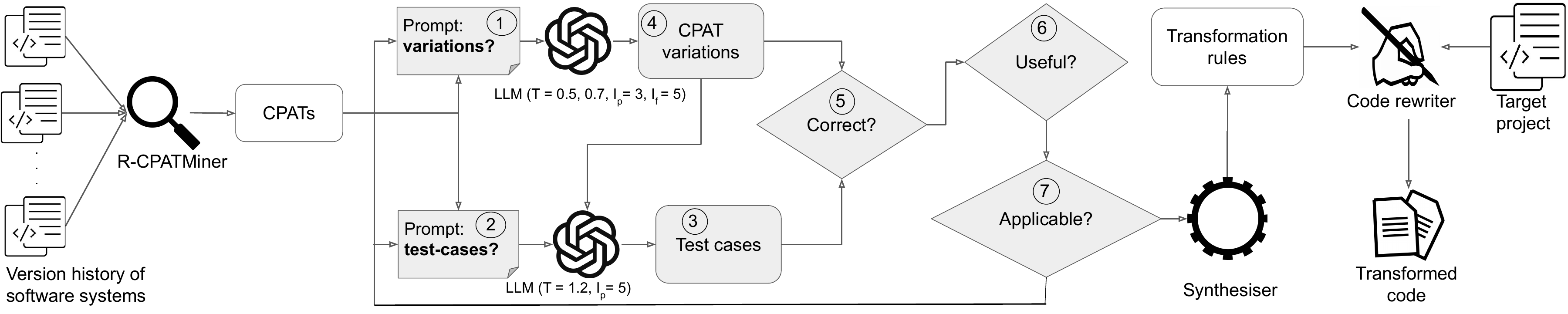}
     \vspace{-4mm}
    \caption{Schematic diagram of \tool}
    \label{fig:workflow}
    \vspace{-3mm}
\end{figure}

In this section, we present how our tool, \toolit, automates unseen variations of \cpats. 
\Cref{fig:workflow} shows the overall architecture of \toolit.
First, to extract \cpats, we use \textit{R-CPATminer}~\cite{rcaptminer2022ICSE}, which has been shown to be highly effective for extracting \cpats from the version history of open-source repositories.
Each \cpat includes multiple code transformations extracted from real-world repositories. 
Then, \toolit takes \cpats as input and invokes an LLM (Step \circled{1} in \Cref{fig:workflow}) to generate unseen variations for original coding idioms involved in the \cpat. 
To ensure correctness, \toolit validates the generated code variations by checking the syntax errors and semantic equivalence against the original \cpat (Step \circled{5} -- detailed in \Cref{sec:correct}). 
Moreover, through the additional fine-tuning parameters of \tool (Step \circled{6}) we generate more useful variations and reduce those that are not useful (detailed in \Cref{sec:useful}).
This step is crucial in identifying realistic variants that are likely to be present in actual code-bases.
The final step involves static code analysis to identify applicable variations (Step \circled{7}) that align with the original \cpat's intent (detailed in \Cref{sec:applicable}). 
Next, \toolit synthesizes a set of transformation rules using both human adaptations mined by \textit{R-CPATminer} and the generated variants. 
Finally, \toolit applies the synthesized transformation rules, generating edits for other target codes.
\tool leverages the core components of \textit{PyEvolve} to infer and apply rules, while \textit{PyEvolve} achieves transformations with precision at \textit{97\%} and a recall of 94\%, ultimately aiding \tool in reliably executing transformations.
These edit suggestions are then presented to developers, who can decide whether to apply them. 
We will present the technical details of each step in the following.

\begin{definition}{(Unseen variant\label{def:unseenvariant})}
 Let $\mathcal{C}$ be a CPAT consisting of code change instances: $C = \{c_1^{LHS} \rightarrow c^{RHS}, c_2^{LHS} \rightarrow c^{RHS}, \ldots\}$. An unseen variant is defined as $v_u^{LHS} \rightarrow v^{RHS}$, where each $c_i^{LHS} \in C$ exhibits either syntactic differences ($\Delta_{\text{syntax}}(v_u^{LHS},$ $c_i^{LHS}) \neq \{\}$) or alterations in data or control flow ($\Delta_{\text{data/control}}(v_u^{LHS},$ $c_i^{LHS}) \neq \{\}$), or both, with respect to $v_u^{LHS}$. 
 The unseen variant ($v_u^{LHS}$) preserves the semantic similarity with the original CPAT.
\end{definition}

\begin{definition}{(Correct variant - $\mathcal{V}_{\text{c}}$\label{def:correctvariant})}
Let $\mathcal{V}$ be a set of unseen variants generated by an LLM for the CPAT $\mathcal{C}$.
A correct variation, denoted as $\mathcal{V}_{\text{c}}$, is a subset of $\mathcal{V}$ that consists of variations semantically equivalent to the original input $c_1^{LHS}$.
Formally, we can express this as:
$\mathcal{V}_{\text{c}} = \{v \in \mathcal{V} \mid S(v, \mathcal{C}) \text{ holds}\}$,
where $S(v, \mathcal{C})$ denotes the semantic equivalence between a variation $v$ and the original CPAT $\mathcal{C}$.
\end{definition}

\begin{definition}{(Useful variant - $\mathcal{V}_{\text{u}}$\label{def:usefulvariant})}
 is a subset of $\mathcal{V}_{\text{c}}$, denoted as $\mathcal{V}_{\text{u}}$, that represents variations that actual programmers would realistically write in their codes.
Formally, we can express this as: 
$\mathcal{V}_{\text{u}} = \{v \in \mathcal{V}_{\text{c}} \mid P(v) \},$ where $P(v)$ represents the condition that a variation  $v$ is considered realistic and likely to be written by programmers.
The term ``not-useful'' is used to describe the correct variants that do not belong to the set of useful variants ($\mathcal{V}_{\text{c}} \setminus \mathcal{V}_{\text{u}}$).
\end{definition}

\begin{definition}{(Applicable variant - $\mathcal{V}_{\text{a}}$\label{def:applicablevariant})}
is a subset of $\mathcal{V}_{\text{u}}$, denoted as $\mathcal{V}_{\text{a}}$, that represents variations with the same structural intent as the original \cpat. 
Formally, this can be expressed as: $\mathcal{V}_{\text{a}} = \{v \in \mathcal{V}_{\text{u}} \mid A(v) \},$ where $A(v)$ represents the condition that a variation $v$ is considered to share the same structural change as the original \cpat. 
\end{definition}

\subsection{Generating Unseen Variations\label{sec:generaingexamples}}
The first step of our approach is to generate a comprehensive set of unseen variations for LHS of each \cpat that belongs to the two types, Variation Type 1: \VTone  and Variation Type 2: \VTtwo.
Here, we consider the LHS (\textit{i.e.}, before code) of the \cpat as the focus for generating these variations since the \cpat specifies that the LHS must be transformed into the RHS (\textit{i.e.}, the after code).

\subsubsection{\textbf{Prepare LLM:}}
LLMs initially trained on extensive datasets may require additional preparation for specific tasks, utilizing in-context learning~\cite{NEURIPS2022_9d560961}.
It provides the LLM with a prompt that prepares it for a particular prediction task. 
As LLMs advance,
in-context learning has evolved into zero-shot learning, enabling predictions solely based on desired output.
However, its application to unexplored tasks remains challenging~\cite{brown2020language, fan2022automated, radford2019language}.
To overcome this challenge, we employ few-shot learning, augmenting the context with a small number of examples representing desired inputs and outputs.
In our work, given that LLMs lack explicit training in understanding, analyzing, and rewriting variants, we utilize few-shot learning~\cite{Gao:2023ASEIn-context} as the in-context paradigm to facilitate LLMs in rewriting variants of \cpats.
We follow best practices as discussed by Gao et al.~\cite{Gao:2023ASEIn-context} and include two examples for \VTone and \VTtwo. 
In addition to examples, the prompt provides LLM with formatting instructions for consistent integration with the tool. 

\subsubsection{\textbf{Optimize Tuning Parameters and Generate  Variations:}\label{sec:tuning_parameters}}
The output of an LLM depends on the internal variable "Temperature" ($T$), that serves as a regulator for the model's output randomness. 
Higher values, such as 0.9 or 1.2, produce more diverse and unpredictable outputs, whereas values closer to 0 produce focused and less diverse results.
Adjusting $T$ allows us to balance between creativity and determinism of the output.  
Furthermore, the output can exhibit variability when the LLM is presented with the same prompt multiple times. 
This because:
\begin{enumerate*}[label=(\roman*)]
\item LLMs generate responses using a combination of learned patterns and random sampling, introducing slight variations in answers each time due to the inherent randomness in the generation process; and 
\item LLMs can explore diverse solutions and improve their responses iteratively through feedback \cite{brown2020language, radford2019language}, enhancing their outputs in subsequent iterations.
\end{enumerate*}
We primarily focus on two types of iterations in this work:
\begin{enumerate*}[label=(\roman*)]
\item prompting the LLM with the exactly same prompt that consists of the same variant ($v_i$), known as prompt iteration ($I_p$), and
\item changing the prompt to augment it with another semantically similar applicable variation generated in a previous step ($v_j$ where $j!=i$), referred to as \fIteration.  ($I_f$).
\end{enumerate*}
While more iterations yield more variations, it also increases processing time and may produce many non-useful variants.
Therefore, selecting the ideal combination of temperature and iteration values is of paramount importance.
Our goal is to fine-tune the parameters \(T\), \(I_p\), and \(I_f\) to maximize the potential for generating diverse and useful variations using the LLMs. 
\Cref{sec:rq3_1} explains our empirical approach to choosing these values.

Not all generated variants are valid for automation. The variations must be correct (\Cref{def:correctvariant}), useful (\Cref{def:usefulvariant}), and applicable (\Cref{def:applicablevariant}). 
In the following sections, we explain how we validate the generated \cpats to ensure they satisfy these requirements.






\subsubsection{\textbf{Selecting Correct Variations\label{sec:correct}:}}

LLM produces many unseen variations for \cpats during the initial step described in \Cref{sec:generaingexamples}. 
Among these variations, it is crucial to identify the correct ones (\Cref{def:correctvariant}) that are syntactically correct and semantically equivalent to the original \cpat, thereby avoiding any erroneous code edit suggestions to the developers.

To ensure the correctness of variations produced by LLMs, we employ a four-step validation process:
\begin{enumerate*}[label=(\roman*)]
\item Syntax validation,
\item Type validation,
\item Import validation, and
\item Semantic validation.
\end{enumerate*}
Syntax validation checks whether the generated variant forms a valid program, devoid of syntax errors such as indentation errors, parentheses errors.
Type validation ensures that variables in both the \cpat and the variant maintain the same type.
For example, the corresponding variable in the generated variant for the variable \smcode{elements} from \Cref{fortonumpy} must be of type \textit{List[int]}.
The import validation step ensures the presence and proper usage of required libraries, modules, or dependencies, verifying that APIs and classes in generated variations originate from the same sources as those used in the original \cpat.
As the variations are partial code fragments, static code analysis alone cannot infer type information or required  imports. 
To address this issue, we utilize LLMs for type and import inference via chain-of-thought reasoning~\cite{NEURIPS2022_9d560961}, which leverages rationales as intermediate steps for LLMs to infer required types and imports. 
This approach facilitates efficient inference of types and imports to evaluate variant conformity with the original \cpats.

The final step, semantic validation, assesses semantic equivalence between generated variations and the original \cpat.
To achieve this, we depend on test cases. However, the original \cpats lack pre-existing test cases, necessitating the generation of new ones.
Automating test case generation is essential for a streamlined pipeline. 
We achieve this by prompting LLMs to generate test cases for the original \cpat.
These test cases validate the semantic equivalence of the generated variations.

To select valid test cases, we follow three steps:
\begin{enumerate*}[label=(\roman*)]
\item We check that the generated test cases are free of syntax errors,
\item We run the test against the original \cpat and disregard the tests that fail. 
This can happen because the generated tests might have invalid assertions, and \item we check that each test initializes every input variable to the \cpat. 
\end{enumerate*}
This last step removes tests that do not instantiate all its variables. 
Such tests may pass the test-case however, using such a test would weed out some correct unseen variants further down the pipeline.
In \Cref{sec:rq2}, we quantitatively study the significance of each of these steps, while
in \Cref{sec:rq3_2}, we analyse the quality of the test cases, ensuring their effectiveness in assessing the semantic equivalence of the generated variations.


\subsubsection{\textbf{Selecting Useful Variations\label{sec:useful}:}}

Useful variations (\Cref{def:usefulvariant}) capture common coding patterns, idioms, or practices that are typically observed in real-world code.
Minimizing the generation of non-useful variants is crucial to limit the inclusion of rules that might not be applicable to the target code.
This is because, when applying the rules to codebases, the code rewriter iterates through all the rules to find matching target codes. 
If there are too many rules that are unlikely to find opportunities for application, it can reduce performance.
However, it is not necessary to completely remove non-useful variations, as they are correct transformations that are semantically equivalent to the original \cpat. 
Furthermore, it is impossible to completely remove them. Therefore, it is acceptable to keep them as variations even if they may not find suitable opportunities for application.
Our goal is to minimize the number of non-useful examples generated by LLMs while increasing the useful ones.
To achieve this, we carefully select parameter values for $T$ and $I$ as explained in \Cref{sec:rq3_1}, by empirically studying its output for varying parameters.


\subsubsection{\textbf{Selecting Applicable Variations\label{sec:applicable}:}} 
Applicable variations, a subset of useful variations, align with the structural intent of the original \cpat. 
Filtering out inapplicable once is crucial to prevent unexpected transformations and ensure that only applicable variations, which preserve desired semantics and adhere to intended transformations, are applied to the code bases.


To identify applicable variants, we assess structural intent using three rules:
\begin{enumerate}[label=(\roman*), wide, labelwidth=!, labelindent=0pt]

\item Control Nodes, as defined by Nguyen et al.~\cite{nguyen2019cpatminer}, are a generalization of AST nodes used to construct fine-grained program-dependence graphs that can be used to group similar code fragments. 
Control nodes are a group of control statements, such as \smcode{if}, \smcode{for}, \smcode{while} statements. 
We first check whether all the control nodes ($ControlNodes(C_{AST}^{LHS})$) in the LHS of the original \cpat are present in the AST of the generated variant $(V_{AST})$. 
Formally, $\forall n \in ControlNodes(C_{AST}^{LHS}) : n \in V_{AST}$.

\item $V_{AST}$ does not contain new declarations, such as method declarations (except variable declarations) that do not exist in the AST of CPAT ($C_{AST}^{LHS}$). 

\item The sign of difference between the AST nodes in the variant and the RHS of the \cpat ($V_{AST} - C^{RHS}_{AST}$) should match the sign of difference of the number of AST nodes between the LHS and the RHS of the \cpat ($C^{LHS}_{AST} - C^{RHS}_{AST}$).
Formally, $(V_{AST} - C^{RHS}_{AST}) \times (C^{LHS}_{AST} - C^{RHS}_{AST}) ~$\textgreater$~0$.

\end{enumerate}

In the first rule, comparing control nodes between the original \cpat's LHS and the generated variant is crucial for determining whether the variant captures the essential control-flow structure intended by the original \cpat. 
The second rule ensures that the variant does not contain unexpected statement declarations, such as new methods or even classes.
In the final rule, by comparing differences in AST node counts, it identifies whether the variant maintains the desired structural transformation, with consistency in signs indicating that the variant introduces or removes the expected AST nodes as \cpat.
These rules play a crucial role in selecting applicable variants as they filter out variants that useful but do not adhere to the desired structural intention of the \cpat.

\subsection{Synthesising Transformation Rules}

TBE techniques infer transformation rules that facilitate code change automation, as exemplified in the code change used to infer the rule.
A complete transformation rule comprises two components: the \textit{Rule} and the \textit{Guard}~\cite{pyevolve2022ICSE,Xiang2021APIFix,Bader2019Getafix}. 
The \textit{Rule} defines specific changes to code fragments, while the \emph{Guard} determines which code the rule should be applied to based on various validations.
For example, in \Cref{lst:ruleNumpySum}, the guard for \smcode{:[[v2]]} is \smcode{type : List[int]}, which verifies whether the corresponding program element in the target code is a list of integers.
There is one or many such validations in a Guard that should be considered when deciding where to apply a rule.

We utilize PyEvolve~\cite{pyevolve2022ICSE} as our rule synthesizer, which leverages InferRule~\cite{inferrule} to infer rules. 
To create a complete code change, we combine each LLM-generated variant with the RHS of the original CPAT and send it to the rule synthesizer.
The rule synthesizer infers a rule for the input code change.
For example, \toolit inferred the rule given in \Cref{lst:newrule} by taking the input code change in \Cref{fortonumpy}, enabling the transformation of the unseen variant shown in \Cref{lst:npsumtype4}.

PyEvolve relies on the output of R-CPATMiner~\cite{rcaptminer2022ICSE} to infer guards for the \cpats extracted from the version history of repositories. 
To infer guards for the program elements in LLM-generated variations, \toolit follows a two-step process:
\begin{enumerate*}[label=(\roman*)]
\item If an element $e$ is in both the original \cpat ($C_{\text{orig}}$) and a variant ($V_a$), then the guard validations for $e$ are inferred from output of \text{R-CPATMiner},
\item if an element $e$ is in a variant ($V_i$) but not in the original \cpat ($C_{\text{orig}}$), then the guard validations for $e$ generated from LLMs.
\end{enumerate*}
Integrating the rule and guard, \toolit infers a comprehensive transformation rule, ensuring systematic and effective code changes.

%% file: Evaluation.tex
\section{Evaluation}

We empirically evaluate \toolit by answering the following research questions:

\begin{enumerate}[label=\bfseries RQ\arabic*.,wide, labelwidth=!, labelindent=0pt]

\item  \textbf{How effective are LLMs at generating variations?}
We depend on LLMs to produce unseen variants, but the ability of LLMs to generate these variants is unknown. 
Therefore, we perform a quantitative analysis of the variants generated by the LLMs with 20 \cpats and utilize the three most recent and largest known LLM models to date.



\item \textbf{How effective are LLMs at generating test-cases?}
While we rely on LLMs to generate test cases for dynamic analysis of the variants, their ability for this task is unknown. 
Therefore, we quantitatively assess the test cases generated by these LLMs across 20 \cpats

\item \textit{\textbf{What are the optimal parameters for generating unseen variants?}}
The randomness and exploratory ability of LLM vary with temperature ($T$) and iteration values ($I$).
Studying how LLMs change their output along with $T$, $I$ values is crucial for enabling \tool to harness the full potential of LLMs. 
To achieve this, we study how quality and quantity of variant generation varies with the Temperature and Iterations values.


\item  \textbf{\textbf{What are the optimal parameters for generating test cases?}}
Tool employs a three-step validation process (\Cref{sec:correct}), to select valid test cases.
To fully harness the potential of LLMs, it is crucial to study how the effectiveness (i.e., quality and quantity) of the test cases varies with the parameters Temperature and Iterations. 


\item  \textbf{How effective is \toolit at finding new opportunities and performing transformations over the baseline?} 
We want to evaluate the improvements over the previous state-of-the-art tools that are solely based on program analysis. 
To achieve this, we performed a comparison with PyEvolve, the leading tool for automating unseen variations. 
Our assessment focused on quantifying the additional opportunities detected and transformed by \toolit in comparison to the baseline.






\item \textbf{How useful are the generated program transformations?}
We want to determine whether developers find code improvements generated by \tool to be useful.
To achieve this, we submit pull requests to open-source projects with the patches generated by \toolit.
\end{enumerate}




\input{RQ1}

\input{RQ2}

\input{RQ3and1}

\input{RQ3and2}

\input{RQ4}

\input{RQ6}

%% file: RQ1.tex
\subsection{RQ1: Effectiveness of Variant Generation\label{sec:rq1}}
Numerous LLMs have been developed, each with its own set of capabilities and applications.
Among them, several of the largest known LLMs include:
\begin{enumerate*}[label=(\roman*)]
\item PALM~\cite{google_bard},
\item GPT-3.5~\cite{brown2020language}, 
and
\item GPT-4~\cite{rae2023gpt}.
\end{enumerate*}
are prominent examples of the largest LLMs developed by Google and OpenAI, boasting trillions of parameters.
The efficacy of \toolit hinges on the choice of the LLM employed. 
Hence, to assess the effectiveness of generating variants using a selected LLM corpus, we have chosen the aforementioned LLM models. 
These models are distinguished by their extensive training on a vast number of tokens, comprising billions, and the significant number of parameters they encompass. 
Furthermore, their internal APIs offer convenient accessibility, an essential factor for seamless integration with \toolit.
We conduct both quantitative and qualitative analyses to evaluate the variants generated by them.

\begin{figure}[t]
    \begin{minipage}[b]{0.48\linewidth}
        \centering
        \includegraphics[height=3.5cm,keepaspectratio=true]{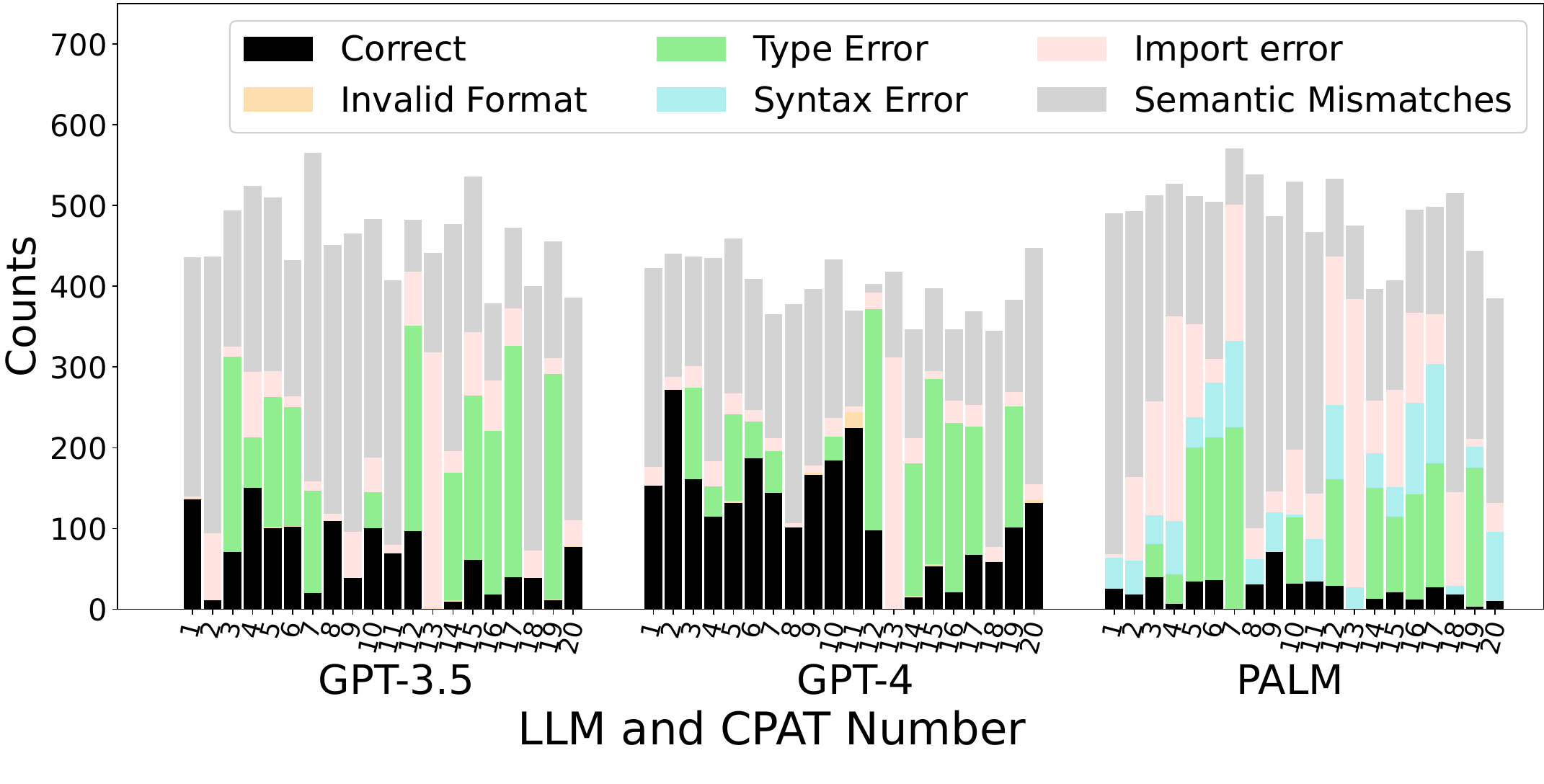}
         \vspace{-4mm}
        \caption{Variation generating capabilities of each LLM}
        \label{fig:variations}
    \end{minipage}
    \hfill
    \begin{minipage}[b]{0.48\linewidth}
        \centering
        \includegraphics[height=3.5cm,keepaspectratio=true]{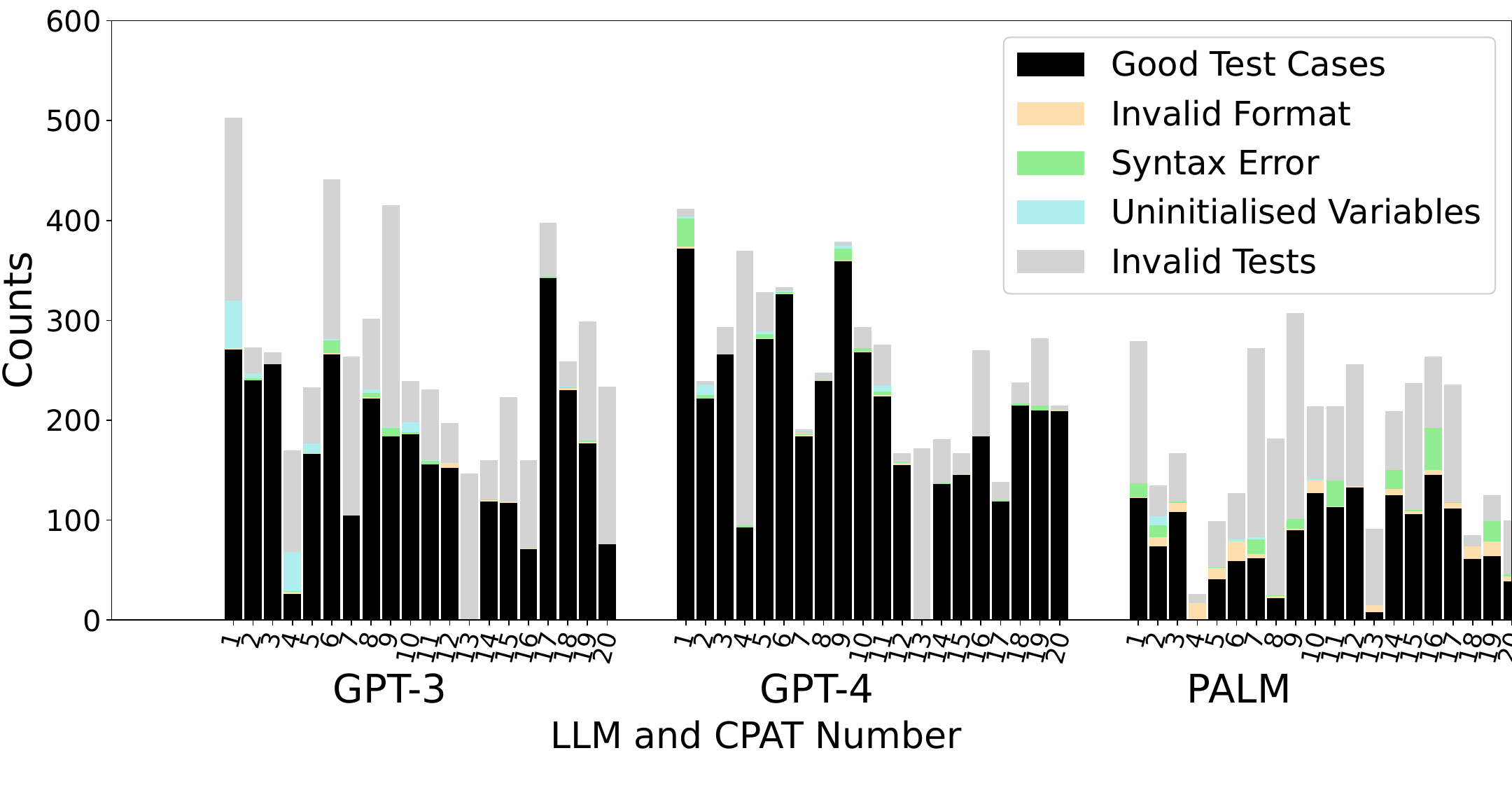}
         \vspace{-4mm}
        \caption{Unit-test generating capabilities of each LLM}
        \label{fig:unittest}
    \end{minipage}
    \vspace{-5mm}
\end{figure}

\subsubsection{\textbf{Dataset}}
Dilhara et al.~\cite{rcaptminer2022ICSE} studied a diverse set of top 2,500 \cpats in Python ML systems, leading to the discovery of four group of frequently occurring \cpats kinds:
\begin{enumerate*}[label=(\roman*)]
\item \trendtwo (e.g., \Cref{fortonumpy}),
\item \trendthree (e.g., \smcode{np.dot(np.dot(A,B),C)} -> \smcode{np.linalg.multi\_dot(A,B,C)}),
\item \trendone (\smcode{open("file.txt")} -> \smcode{with open ("file.txt")}), and
\item \trendfour (e.g., Python list comprehension).
\end{enumerate*}
The authors' survey involving 650 developers further confirmed the developers' strong interest in automating these identified \cpats across all four categories. 
Thus, we select a representative sample of 20 \cpats covering all four kinds to comprehensively answer this research question.

\subsubsection{\textbf{Experimental Setup}}
We initiated each LLM with the prompt for each of the 20 \cpats. 
This generated many raw variants for each CPAT. 
Then, we executed \toolit's validation components to thoroughly examine the four-step validation process described in \Cref{sec:correct}, which includes:
\begin{enumerate*}[label=(\roman*)]
\item Syntax validation, 
\item Type validation,
\item Import validation, and
\item Semantic validation.
\end{enumerate*}
For each \cpat, we provide a breakdown of variants, showing the count of instances failing each validation.
Variants that successfully pass all validations are categorized as correct and are suitable candidates for further processing.
This analysis helps understand LLMs' variant generation capabilities and underscores the importance of static analysis and validation before selecting variants for further processing.

\subsubsection{\textbf{Results}}


Our initial observations indicated that PALM, GPT-3 and GPT-4 consistently produced a wide array of variants for all \cpats. 
\Cref{fig:variations} shows the quantity of variants generated by each LLM for every CPAT. 
Each bar within the chart represents distinct variants, categorized as those with syntax errors, type errors, import errors, semantic mismatches, and, finally, correct variants suitable for subsequent processing.
We noticed that all LLMs consistently generated numerous raw variations for all the evaluated \cpats.
PALM excelled compared to other LLMs in producing a greater quantity of raw variants, with a notable achievement of 584 variants—surpassing the highest count generated by any LLM for a single CPAT. 
On average, it generated 518 variants per CPAT, outperforming GPT-4, which produced 398 variants, and GPT-3.5, which yielded 461 variants.
This highlights the remarkable creativity and capabilities of LLMs in generating many variations.

However, it is worth noting that the relative performance of LLMs varies significantly in terms of generating correct variations.
Notably, PALM consistently exhibited a propensity for generating a higher total number of variants across all \cpats compared to GPT-3.5 and GPT-4. However, on average, only 5\% of these variants remained free from errors and proved to be correct.
Conversely, GPT-4, while producing fewer raw variants than PALM and GPT-3, consistently generating a greater number of distinct correct variants than all LLMs.
Quantitatively, on average, GPT-4 generated 102\% and 555\% more correct variants than GPT-3.5 and PALM, respectively.
Despite GPT-4's ability to generate the highest number of correct variants, it still produced incorrect variants at a average rate of \LLMerrorrate. This underscores the imperative need for filtering techniques, aka ``trust but verify''.

\resultbox{
LLMs excel in generating unseen variants but also produce errors. GPT-4 generates the most correct variants, but its \LLMerrorrate error rate emphasizes the need for error identification techniques.}



%% file: RQ2.tex
\subsection{RQ2: Effectiveness On Test Case Generation\label{sec:rq2}}
\toolit utilizes test cases generated by LLM to verify the semantic equivalence between the original input \cpat and the generated variations.
Therefore, we evaluate both the LLM's proficiency in generating test cases and the quality of these generated test cases. 
To achieve this, we conduct a two-fold analysis: initially, a quantitative assessment of LLM's test case generation capabilities, followed by an investigation into the quality of these test cases using mutation testing techniques. 
Mutation testing involves deliberately introducing subtle modifications (mutations) to the \cpat and subsequently retesting the generated tests to determine their ability to detect these mutations. 
This technique is instrumental in identifying tests that may not effectively uncover faults.

\subsubsection{\textbf{Dataset and Experimental Setup}\label{sec:rq3_data}} 
We chose the LLMs detailed in \Cref{sec:rq1} and prompted them to generate test cases for all 20 \cpats considered in \Cref{sec:rq1}.
Before selecting the test cases that will help us choose the correct variants, we followed a three-step validation process, elaborated in Section \ref{sec:correct}. 
This process ensured that the test cases meet the following criteria:
\begin{enumerate*}[label=(\roman*)]
\item they are devoid of syntax errors,
\item they initialize all its variables, and
\item original \cpat pass the test case.
\end{enumerate*}
Our quantitative analysis is primarily focused towards evaluating test cases based on these steps, identifying those that do not conform to the criteria as erroneous test cases.
Furthermore, we conducted mutation testing on the test cases that successfully met all steps.
This aids in comprehending the efficacy of the test cases intended for the selection of correct variants.
To generate mutants, we used the widely used \emph{mutmut}~\cite{mutmut}, known for generating Python mutations.

\subsubsection{\textbf{Results}}
Our observations revealed that PALM, GPT-3, and GPT-4 consistently produced output with test cases for all the input \cpats.
As shown in Figure \ref{fig:unittest}, we observed that LLMs are capable of generating many unit tests for a given code idiom. 
In certain cases, it produced up to 503 test cases with GPT-3, resulting in 271 valid test cases among them. 
GPT-3 outperformed all the LLMs in terms of generating a higher number of test cases. However, it produced test cases with errors, described in \Cref{sec:rq3_data}, at an average rate of 37\%. 
In contrast, GPT-4 generated a relatively lower number of test cases compared to GPT-3 but still more than PALM, and its error rate was significantly lower, standing at 19\%. 
Therefore, GPT-4 produced the highest number of error-free test cases. 
We observed that GPT-4, on average, created 210 error-free test cases per \cpat, ranging from 93 to 372, demonstrating the LLM's capability to generate extensive test suites. This also underscores the importance of employing techniques to select error-free test cases.

We then executed the tools to generate mutants of the \cpats and employed the generated test cases to detect both the mutants and the original code. Our observations revealed that the generated test cases achieved a 100\% success rate when detecting the mutants generated by the mutation testing tool, \textit{mutmut}~\cite{mutmut}. 
In \Cref{sec:rq3_2}, we further analyze the effectiveness of these generated tests by comparing their performance to an oracle of human-generated test cases.

\resultbox{
LLMs excel in test case generation but also introduce errors. While GPT-4 outperforms others, its 19\% erroneous test cases emphasize the importance of validation techniques.
}






%% file: RQ3and1.tex
\subsection{RQ3: Best Performing Parameters for Generating Variants\label{sec:rq3_1}}

\input{images/table-evaluations}

The LLM's output varies with temperature ($T$), prompt iteration ($I_p$), and \fIteration ($I_f$), as they impact randomness and iterative exploration, leading to variations in the generated responses (see \Cref{sec:tuning_parameters} for parameter details).
Therefore, we adopt an empirical approach to determine optimal values for generating variants.
We first generated an oracle of unseen variants consist with 
9325 unseen variants for 10 \cpats given \Cref{table:cpat}. 
Then, we used both automated and manual steps to group the variants into \textit{Correct variants} (\Cref{def:correctvariant}), \textit{Useful variants} (\Cref{def:usefulvariant}), and \textit{Applicable variants} (\Cref{def:applicablevariant}).
Then, we study the generation of these variants in relation to the parameters to determine the best-performing settings for \tool.



\subsubsection{\textbf{Creating an Oracle:}\label{sec:oracal}}
\begin{figure}[t]
    \begin{minipage}[b]{0.5\linewidth}
        \centering
\includegraphics[height=2cm,keepaspectratio=true]{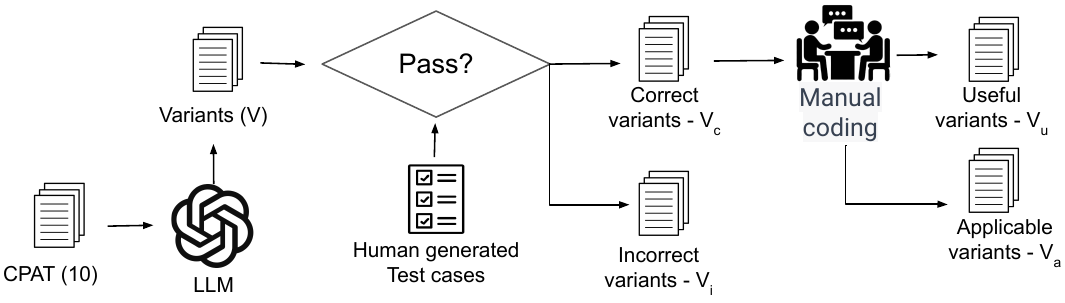}
         \vspace{-5mm}
    \caption{Schematic diagram of the workflow for generating data to fine-tune LLM parameters}
    \label{fig:empirical}
    \end{minipage}
    \hfill
    \begin{minipage}[b]{0.38\linewidth}
        \centering
        \includegraphics[width=5cm,keepaspectratio=true]{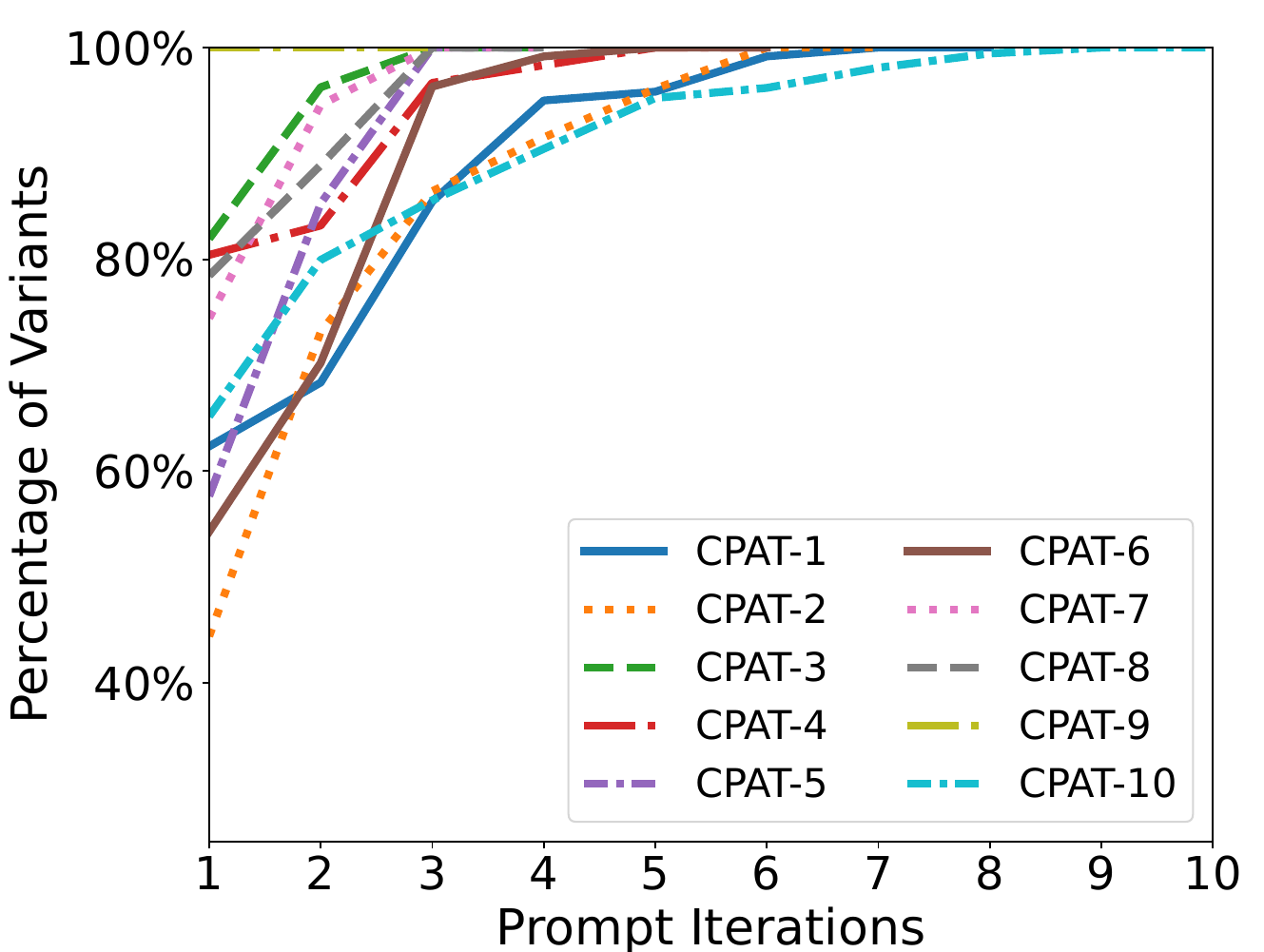}
         \vspace{-4mm}
        \caption{Generation of variants (Y-axis) along with prompt iteration (X-axis)}
        \label{fig:promptIteration}
    \end{minipage}
    \vspace{-5mm}
\end{figure}

As shown in \Cref{fig:empirical}, we invoke LLM to generate unseen variants for the 10 CPATs
shown in \Cref{table:cpat}.
Our objective is to generate a comprehensive set of all possible unseen variants for each CPAT.
We generate 10 such sets, and each set is denoted as $V^m$, where $m$ corresponds to the CPAT number ($m \in \{1, 2, \ldots, 10\}$).
To create $V^m$, we prompted LLM with \cpat $m$ for each temperature value in the set $\{0, 0.3, 0.5, 0.7, 0.9, 1.2\}$.
For each temperature value, we performed prompt iterations ($I_p$) up to 15 times, by repeatedly prompting the LLM with the same CPAT.
After completing all 15 prompt iterations, we employed \fIteration, inputting each generated variant back into the LLM. We repeated these steps until no new variants were generated.
Through this process, a total of 8064 distinct variants were generated for all CPATs.

As the next step, we statically validated the generated variants, as described in \Cref{sec:correct}.
Further, two authors of the paper manually wrote unit test cases for the \cpats to evaluate their functionality and consider boundary cases. 
Then, we executed the test cases on the generated unseen variants to classify them as either correct ($V_c^m$) or incorrect ($V_i^m$). 
This classification process resulted in total \rqfourVI incorrect variants ($\sum_{m=1}^{10}$ $V^m_i$) and \rqfourVC correct variants ($\sum_{m=1}^{10}$ $V^m_c$).

To categorize the correct variants ($V^m_c$) into "Usable" (\Cref{def:usefulvariant}), referred to as ($V^m_u$), we utilized the Inter-Rater Reliability (IRR) methodology~\cite{landis1977measurement}. Following best practices and qualitative research guidelines, the authors employed a negotiated agreement technique~\cite{wicks2017coding, campbell2013coding} to achieve consensus.
Two authors of the paper conducted manual analyses on each change pattern to identify the high-level programming tasks associated with them. They reached a consensus to classify a variant as ``usable'' if it represented something a developer would realistically write, and they labeled variants as ``not-useful'' if developers would not typically include them in their code.
Then, we conducted the static analysis given in \Cref{sec:applicable} to further identify the ``applicable'' variations ($V^m_a$) from the ``useful'' variants. With this process, we identified a total of 1039 useful variants ($\sum_{m=1}^{10}$ $V^m_u$) and 580 applicable variants ($\sum_{m=1}^{10}$ $V^m_a$).
This data set serves as a resource for fine-tuning variables during the generation of test cases and variants as explained in the following sections.


\noindent
\subsubsection{\textbf{Prompt Iteration $(I_p)$}}
We study the number of prompt iterations required for each temperature value until it no longer produces a significant number of new variants, which guides our decision-making regarding the suitable number of prompt iterations for subsequent steps.

\noindent
\textbf{Dataset and Experimental setup:}
We input the CPATs from \Cref{table:cpat} into the LLM alongside a prompt for variant generation, iterating with the same prompts labeled as $i^x_p$ from the 1st to the 100th iteration, where $x$ represents the iteration number.
Then, we study the cumulative count of distinct variants produced during each iteration.

\noindent
\textbf{Results:} 
\Cref{fig:promptIteration} shows the process of variant generation for the initial 10 iterations at temperature 0.
In the first iteration, the LLM generated an average of 70\% (with a minimum of 45\% and a maximum of 82\%) of total variants. 
By the second iteration, this increased to 83\% (with a minimum of 68\% and a maximum of 96\%) of variants. 
During the third iteration, the LLM generated 95\% (with a minimum of 85\% and a maximum of 100\%) of variants. 
Beyond that point, we observed that the LLM hardly generates any new variants for a given \cpat.

{To statistically analyze the data, we started with forming distributions as $D^t_i = \{d_x \mid d_x$ \textit{represents the proportion of total variants generated at the $t^{\text{th}} temprature,$ by the completion of the } $i^{\text{th}} $\textit{prompt iteration} for the  $x^{\text{th}} $\text{ \cpat}, $\forall x \in \{1, 2, \ldots, n\}\}$.
We then used the \textit{Wilcoxon Signed-Rank Test} to analyze the paired samples $(D^t_i, D^t_{i+1})$, $\forall t$. This analysis focused on iterations $i$ within the set $\{1, 2, 3, 4, 5\} \subset \mathbb{N}$, which were marked by significant variant generation. The null hypothesis, positing no significant difference between the variant observations in pairs $(D^t_i, D^t_{i+1})$ for all $i$ and $t$, was rejected for $i<3$ and $t\neq0$. 
To quantify the differences between these distributions, we utilized the \textit{Hodges-Lehman estimator}, revealing differences as follows: for $i \in \{1, 2, 3, 4\}$, $D^t_{i+1} - D^t_i = \{25\%, 16\%, 0\%, 0\%\}$ respectively where $t = 0.5$, indicating a potential for increased variant generation with each iteration up to the third iteration. 
However, this incremental benefit was noted to diminish after the third iteration. 
This observation is consistent for all $t$ except $t=0$. 
At $t=0$, the randomness is reduced to zero, resulting in the LLM conservatively generating a limited number of variants, all of which are produced at $i=0$ without extending beyond that point.
Consequently, to optimize the trade-off between the merits of additional variant generation and the implications of extended processing duration, we opted to cap the iteration count at three.}





\resultbox{Continuing prompt iterations indefinitely is an option to generate numerous variants. However, beyond the third iteration, LLM produces new variants sparingly.}

\subsubsection{\textbf{\FIteration $(I_f)$:}}
In order to generate as many variants as possible, we perform \fIteration by inputting an applicable variant generated in the previous step to the LLM. 
While this approach can potentially generate many variants, it can also lead to producing numerous not-useful variants (refer \Cref{sec:useful}) when the LLM is prompted with another not-useful variant. 
This situation can impact the performance of \tool in two ways: 
\begin{enumerate*}[label=(\roman*)]
\item multiple iterations generating not-useful variants, 
and 
\item inferring rules for not-useful variants that will not be applied to the code.
\end{enumerate*}
We study the generation of useful variants along with the number of \fIterations.

\textbf{Dataset and Experimental setup:} 
We use the oracle mentioned in \Cref{sec:oracal} to study this in two ways. 
\begin{enumerate*}[label=(\roman*)]
\item We study, for each temperature, the cumulative ratio of non-useful variants to total non-useful variants in the oracle, generated at each \fIteration.
This analysis helps us understand the trends in generating non-useful variants along with \fIteration.
\item In addition to minimizing non-useful variants, it is also important to increase the number of useful cases. 
Hence, we study the generation of useful cases along with iterations and temperature.
\end{enumerate*}

\textbf{Results:} 
The plot in \Cref{fig:hillusinations} shows the distribution of the ratio of the cumulative number of non-useful variants generated by each iteration relative to the total non-useful variants in the oracle.
This analysis is conducted for each \cpat listed in \Cref{table:cpat}, represented by lines in each plot, and across various temperature values.
While \Cref{fig:hillusinations} presents the plot for only temperature 0, the others showcase similar trends as depicted in the provided plots and are available on our companion website~\cite{artifacts}.
We noticed a consistent trend across all temperatures and each \cpat, where the generation of non-useful variants begins after approximately 3 to 6 \fIterations, followed by a rapid increase.

{To statistically analyze the data, we define distributions as:
$F^t_i = \{f_x \mid f_x$ \textit{ represents the proportion of total non-useful variants generated at the } $t^{\text{th}}$ \textit{ temperature by the completion of the } $i^{\text{th}}$ \textit{\fIteration for the } $x^{\text{th}} \cpat, \forall x \in \{1, 2, \ldots, n\}\}.$
We then applied the Wilcoxon Signed-Rank Test to analyze the paired samples $(F^t_i, F^t_{i+1})$, $\forall t$. This analysis focused on iterations $i$ within the set $\{2, 3, 4, 5\} \subset \mathbb{N}$, identified as the iterations where significant generation of non-useful variants begins. The test consistently rejected the null hypothesis, suggesting a significant difference between the variant observations in pairs $(F^t_i, F^t_{i+1})$ for all $i$ and $t$. To quantify these differences, we employed the Hodges-Lehman estimator, which revealed that for $i \in \{2, 3, 4, 5\}$ and $t = 0.5$, the differences are:
$F^t_{i+1} - F^t_i = \{1\%, 2.5\%, 3.6\%, 5.1\%\}$
respectively, indicating an increasing trend in non-useful variant generation starting significantly after the third iteration.}
{Furthermore, to understand whether the production of non-useful variants behaves the same across all temperature values, we applied again the Wilcoxon Signed-Rank Test to analyze the paired samples for all \((F^{t_1}_i, F^{t_2}_i)\), where \(t_1 \neq t_2\). At each \fIteration \(i\), no significant differences were observed in the distributions, indicating that temperature values behave consistently in producing non-useful variants.}


While we observe that reducing non-useful variants is more achievable with fewer \fIterations, a trade-off arises because we also aim to increase the generation of useful variants. 
To address this, we analyzed the cumulative ratio of useful variants generated up to each iteration compared to the total variants generated, across different temperature values.
Consistently across all \cpats, we observed that temperatures within the middle range, i.e, 0.5 and 0.7, consistently yield useful variants with fewer \fIterations compared to other temperatures.
For example, with four \fIterations, temperatures 0.5 and 0.7 yield useful variants spanning 61\% to 83\% for each \cpat; this percentage rises to 69\% to 88\% with five iterations, and the trend persists with further iterations.  
{To statistically analyze the data, we defined distributions as: \(U^i_t = \{u_x \mid u_x\) represents the proportion of total useful variants generated at the \(t^{\text{th}}\) temperature, from the temperature list \( \text{temp} = \{0, 0.3, 0.5, 0.7, 0.9, 1.2\}\) relative to the total useful variants in the oracle up to the \(i^{\text{th}}\) \fIteration for the \(x^{\text{th}}\) \cpat, for all \(x \in \{1, 2, \ldots, n\}\}\). Then, we applied the Wilcoxon Signed-Rank Test to analyze the paired samples \((U^i_{t1}, U^i_{t2})\), where \(t1 \neq t2\) for all \(i\), to check for statistically significant differences. This was followed by the Hodges-Lehman estimator to compute the differences in useful variant generation between each temperature value, after which they were ranked according to the estimator. The tests found that temperatures \(0.5\) and \(0.7\) produced a higher number of useful variants compared to all other temperature values.}

Our goal is to generate more useful variants while minimizing non-useful ones. We determined that performing five \fIterations and focusing on variations generated at middle temperatures of 0.5 and 0.7 is an optimal approach. Using these settings, we further apply \tool in \Cref{sec:rq4} and \Cref{sec:rq6} to compare its effectiveness against the baseline. Subsequently, we present the generated variants to actual developers for usefulness evaluation.

\resultbox{Medium temperatures (0.5 - 0.7) yield fewer non-useful variants and more useful variants, all while requiring fewer \fIterations.}

%% file: images/table-evaluations.tex
\begin{table*}[t]
	\centering
	\caption{Variant generation for each \cpats, and the number of transformations.}
	\vspace{-2mm}
\fontsize{5pt}{6pt}\selectfont
	\label{table:cpat}
	\begin{threeparttable}	
		\begin{tabular}{
		l@{\hspace{0.5\tabcolsep}}|
		@{\hspace{0.5\tabcolsep}}
		l@{\hspace{0.5\tabcolsep}}
		l@{\hspace{0.5\tabcolsep}}
         l@{\hspace{0.5\tabcolsep}}|
		l@{\hspace{0.5\tabcolsep}}|
		l@{\hspace{0.5\tabcolsep}} |
		l@{\hspace{0.5\tabcolsep}}|
		l@{\hspace{0.5\tabcolsep}}|
  		l@{\hspace{0.5\tabcolsep}}|
        l@{\hspace{0.5\tabcolsep}}}
        
            \toprule
\# & LHS of CPAT  &RHS of CPAT &  &\textbf{V} &\textbf{$V_c$}   & \textbf{$V_u$} & \textbf{$V_a$} &$T_{1}$ & $T_{2}$ \\\midrule

1
&
{\begin{lstlisting}[style=tablecode, language=python]
count = 0
for i in int_list:
   count =i + count
\end{lstlisting}}
&
\begin{lstlisting}[style=tablecode, language=python]
count = np.sum(int_list)
\end{lstlisting}
&

&1185 &291 &83 & 50  & 17 &  196 (11x) \\
 \hline

2
& 
\begin{lstlisting}[style=tablecode, language=python] 
for k, v in add_dict.items():
    d[k] = v 
\end{lstlisting}
&
\begin{lstlisting}[style=tablecode, language=python]
d.update(add_dict)
\end{lstlisting}
&
&1201 &478 & 119& 110  & 51 & 201 (4x) \\
 \hline

3
&

\begin{lstlisting}[style=tablecode, language=python] 
common = []
for i in l1:
    if i in l2 and i not in common:
        common.append(i)
\end{lstlisting}
&
\begin{lstlisting}[style=tablecode, language=python] 
common = list(set(l1).
            intersection(l2))
\end{lstlisting}
&

&782 & 287 & 107 &66 & 10

 &141 (14x)  \\
 \hline
4
&
\begin{lstlisting}[style=tablecode, language=python] 
for idx, item in enumerate(values):
    if idx != 0:
        string += ", "
    string += item
\end{lstlisting}
& 
\begin{lstlisting}[style=tablecode, language=python] 
string = +", ".join(values)
\end{lstlisting}
&
&285 &101 & 20& 10 & 2 & 12 (6x)
\\

\hline

5
&
\begin{lstlisting}[style=tablecode, language=python] 
d = {}
for i in array:
  if i in d:
    d[i].append(f(i))
  else:
    d[i] = [f(i)]
\end{lstlisting}
& 
\begin{lstlisting}[style=tablecode, language=python] 
d = {}
for i in array:
    d.setdefault(i, []).append(f(i))
\end{lstlisting}
&

&1265 &416 & 150& 75& 9& 125 (14x) \\
\hline
6
&
\begin{lstlisting}[style=tablecode, language=python] 
counts = {}
for i in iterable:
    if i not in counts:
        counts[i] = 0
    counts[i] += 1
\end{lstlisting}
& 
\begin{lstlisting}[style=tablecode, language=python] 
counts = Counter(iterable)
\end{lstlisting}
&
&927 & 425& 202 & 85 & 11 & 
106 (10x)

\\
\hline
7
&
\begin{lstlisting}[style=tablecode, language=python] 
cum_arr = []
for i in range(len(array)):
    cum_arr.append(sum(array[:i+1]))
\end{lstlisting}
& 
\begin{lstlisting}[style=tablecode, language=python] 
cum_arr = np.cumsum(array)
\end{lstlisting}
&
& 1223&290 &95 & 80&3 & 68 (23x) \\

\hline
8
&
\begin{lstlisting}[style=tablecode, language=python] 
dot_prod = 0
for i in range(len(arr1)):
    dot_prod += arr1[i] * arr2[i]
\end{lstlisting}
& 
\begin{lstlisting}[style=tablecode, language=python] 
dot_prod = np.dot(arr1, arr2)
\end{lstlisting}
&
&177 &28 &26 & 24& 16 & 208 (13x)\\

\hline
9
&

\begin{lstlisting}[style=tablecode, language=python] 
result = []
for i in range(len(array1)):
    result.append(array1[i] + array2[i])
\end{lstlisting}
& 
\begin{lstlisting}[style=tablecode, language=python] 
result = np.add(array1, array2)
\end{lstlisting}
&
&64 & 11 &11 & 9& 5& 
36 (7x) \\
 \hline
10
&
\begin{lstlisting}[style=tablecode, language=python]
t = []
for i in range(len(elem)):
    if cond(elem[i]):
         t.append(elem[i])           
\end{lstlisting}

&
\begin{lstlisting}[style=tablecode, language=python]
t = [elem[i] for i in range(len(elem)) 
        if cond(elem[i])]
\end{lstlisting}
&

&955 &453 &226 & 71 &23 & 907 (39x)  \\


\midrule



		\end{tabular}
		\begin{tablenotes}[para]
        \item \textbf{V}: Number of unseen variations generated by LLM,
        \item \textbf{$V_c$}: Number of correct variations,
        \item \textbf{$V_u$}: Number of useful variations,
        \item \textbf{$V_a$}: Number of applicable variations,
        \item \textbf{$T_1$}: Number of application performed by PyEvolve, 
        \item \textbf{$T_2$}: Number of applications performed by \tool.
        \end{tablenotes}

	\end{threeparttable}
\end{table*}

%% file: RQ3and2.tex
\subsection{RQ4: Best Performing Parameters for Generating Test Cases\label{sec:rq3_2}}
\tool generates test cases to identify semantically equivalent variants for the original \cpat. 
We employ LLM with varying temperatures ($T$) and repeated prompt iterations ($I_p$) to generate multiple test cases for a single \cpat, fostering diversity through the influence of randomness and iterative exploration.
We use the oracle generated in \Cref{sec:oracal} to study the effectiveness of test cases produced by \tool. 
We then determine the optimal combination of values for $T$ and $I_p$to generate effective test cases.



\begin{figure}[t]
        \begin{minipage}[b]{0.35\linewidth}
        \centering
        \includegraphics[width=0.9\linewidth,keepaspectratio]{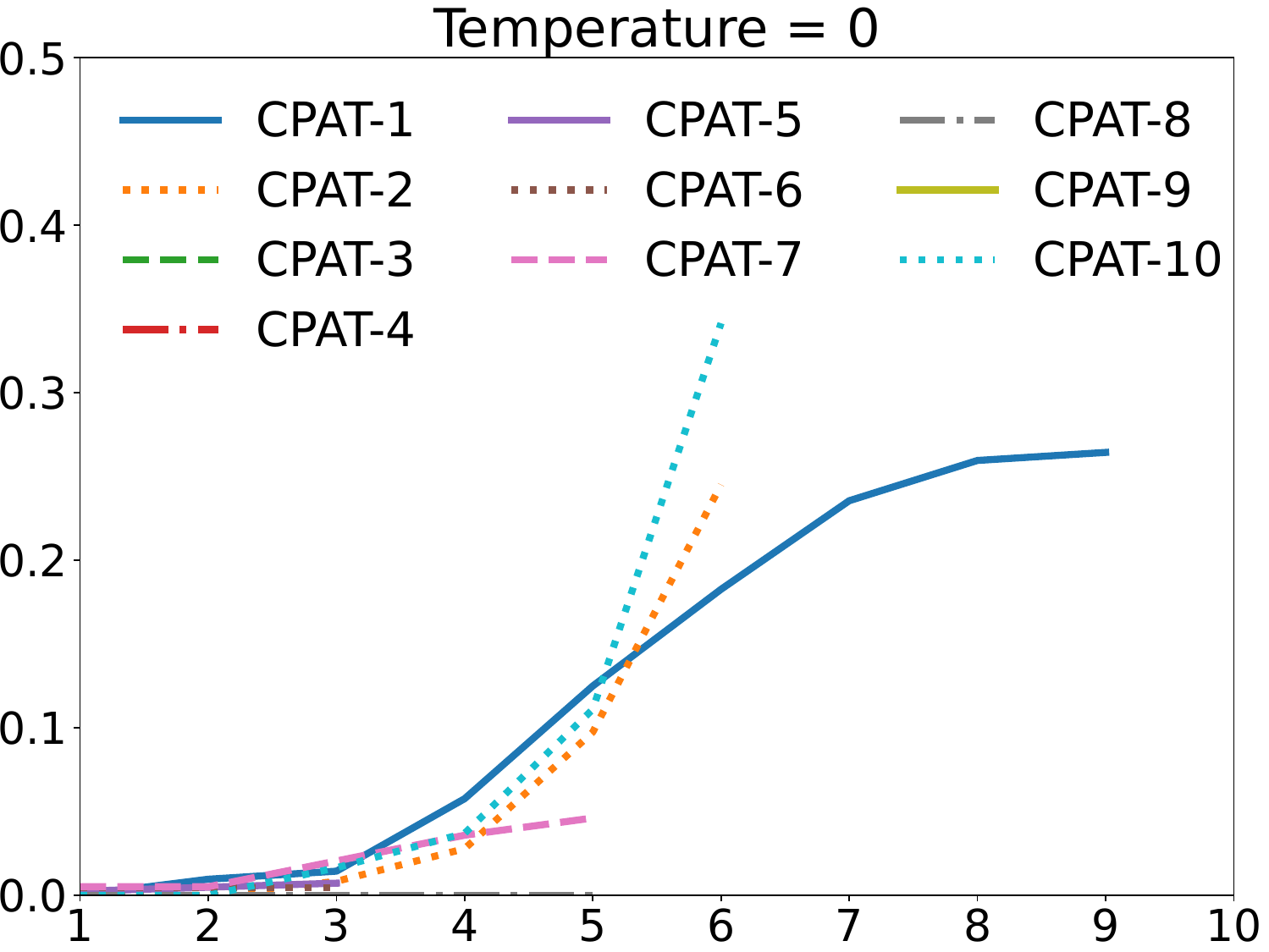}
         \vspace{-3mm}
        \caption{Generation of not-useful variants along with \fIterations}
        \label{fig:hillusinations}
    \end{minipage}
    \hfill
    \begin{minipage}[b]{0.62\linewidth}
        \centering
        \subfigure{
        \includegraphics[width=0.75\linewidth]{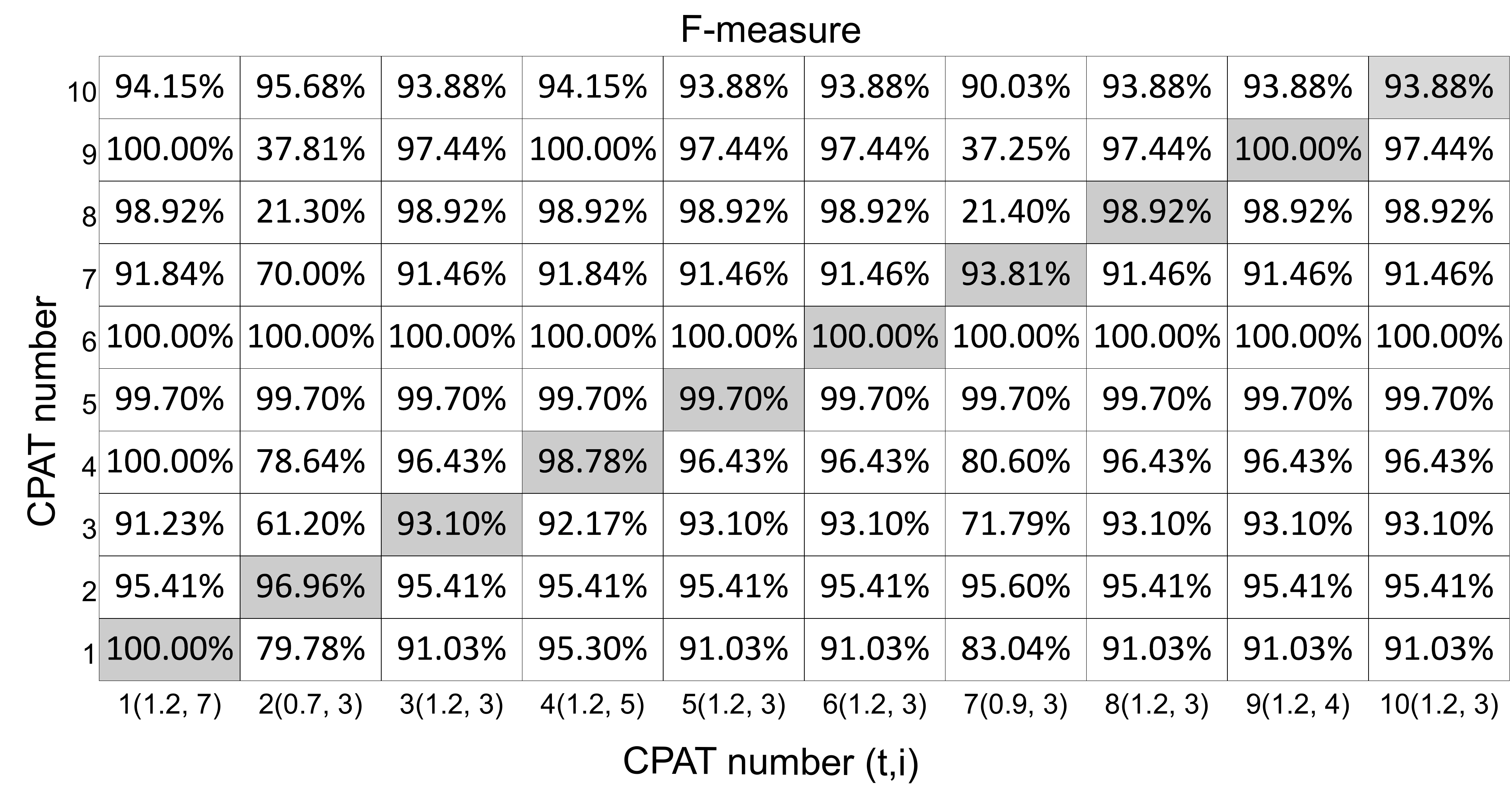}
        }
        \vspace{-5mm}
        \caption{Cross-validation results for evaluating optimal $i$ and $t$ \\ values to generate test cases\label{fig:metrics}}
    \end{minipage}
    \vspace{-4mm}
\end{figure}

\subsection{Dataset and Experimental Setup:}
To study the optimal parameters for test case generation, we study the quality of the generated test cases for a range of Temperature (\(T\)) values, from the set \(\{0, 0.2, 0.4, 0.6, 0.8, 1.0, 1.2\}\), and prompt iterations (\(I_p\) = \(\{1,2, \dots, 15\}\)).
We generate a set of test cases using LLM for each \cpat, considering each \((t, i)\) combination where \(t \in T\) and \(i \in I_p\). 
These generated test cases are then subjected to the three-step validation process utilized by \tool.
This process enables us to identify and select the valid test cases, denoted as \(T^m_{(t,i)}\), across diverse \(T\) and \(I_p\) values.
We then study the effectiveness of testcases, by
executing each set of test cases, $T^m_{(t,i)}$, on the oracle $V^m$ (described in \Cref{sec:oracal}) and classified them into correct variants $\overline{V}^m_c$ and incorrect variants $\overline{V}^m_i$, and then compare them to the classified variants \(V^m_i\) and \(V^m_c\) in the oracle.
To evaluate the effectiveness of generated test cases in classifying variants as correct or incorrect, we calculated precision and recall. 
Precision is the percentage of correct variants (\(\overline{V}^m_c \cap V^m_c\)) among the variants classified as correct (\(\overline{V}^m_c\)), while recall is the percentage of correctly identified variants among all actual correct variants in the oracle (\(V^m_c\)).
Then, we compute F-measure using precision and recall values which is the harmonic mean of precision and recall, calculated as \(\frac{2 \cdot (\text{precision} \cdot \text{recall})}{\text{precision} + \text{recall}}\).

To identify the most effective combinations of \(i\) and \(t\), we adopt the following approach: For each fixed value of \(i\), we calculate the F-measure for all \(t\) values. 
Then, for each \(i\), we identify the \(t\) value that yields the maximum F-measure, denoted as \(\max_{t \in T} F(t, i)\), along with the corresponding \(t\) value, denoted as \(t_{\text{max}}\). 
To further refine this process, we introduce a convergence criterion. If the difference between the maximum F-measure of the current iteration, \(F(t_{\text{max}}, i)\), and the maximum F-measure from the previous iteration, \(F(t_{\text{max}}, i-1)\), exceeds a threshold \(\delta\), we increment \(i\) by 1 and repeat the procedure. 
This iterative refinement goes until the F-measure no longer improves by more than \(\delta\).
We have chosen \(\delta\) to be 5\% to ensure swift convergence based on the observed data.

After selecting a specific $(t, i)$ combination for $V^p$ that effectively generates test cases to classify the variants, we perform 10-fold cross-validation to assess whether this chosen combination learned from one \cpat can be generalized to other CPATs. 
Let $V^m$ be a set of variations generated for a CPAT, where $m \in {1, 2, 3, \ldots, 10}$. 
After obtaining the optimal combination $(t, i)$ for $V^m$, we then apply this learned combination to all the other $V^{m'}$s, where $m' \in {1, 2, 3, \ldots, 10}$ and $m' \neq m$. 
This allows us to assess how well this combination generalizes to different patterns.

\subsection{\textbf{Results:}}
\Cref{fig:metrics} shows the cross-validation results. 
The x-axis represents the $(t, i)$ configurations that achieved F-measure values for each CPAT index (refer to \Cref{table:cpat} for index to CPAT mapping), while the y-axis represents CPAT index.
The outcomes of k-fold cross-validation involve applying the learned \((t, i)\) values to generate test cases with \tool for other patterns and then calculating their precision and recall values, which are displayed in the respective cells.

The analysis consistently shows that $t = 1.2$ yields higher F-measure, indicating that LLM with a higher degree of randomness is better suited for test case generation. Additionally, the number of iterations varies between $3$ and $10$ for each CPAT, with higher iterations generally resulting in similar or better precision and recall. However, to strike a balance between effectiveness and efficiency, we choose the value for $i = 5$, which provides superior precision and recall for \cpats. 
The combination of $(t, i)$ strikes a balance between precision, recall, and efficiency, making it the ideal setting for test case generation by \tool with F-measure 96.6\%.

{To statistically analyze this, we use the computed F-measure values, $F^i_{\text{temp[t]}}$, for each pattern when detecting correct variants under all the temperature and iteration settings. We formed the distribution, $F^i_{\text{temp}}[t] = \{f_m \mid f_m$ \textit{represents the F1 measure of detecting correct variants for pattern} $m^{th}$ \cpat \textit{where} $m \in \{1, 2, \ldots, 10\}$ at the $i^{th}$ \textit{prompt iteration} ($i \in \{1, 2, \ldots, 10\}$) \textit{and at the} $t^{th}$ \textit{temperature setting from the temperature list} $\text{temp} = \{0, 0.3, 0.5, 0.7, 0.9, 1.2\}\}$. We formed 60 distributions for each $i, t$ setting, then selected two distributions at a time and performed the Wilcoxon Signed-Rank Test to analyze the paired samples of F1 scores for each pattern in the two distributions. The test rejected the null hypothesis, indicating a significant difference between the F1 scores across distributions. Subsequently, we applied the Hodges-Lehman estimator to each combination of distributions to quantify the difference in F1 score for each $i$ and $t$ setting. Distributions were then ranked based on the quantifier, revealing that the F1 scores computed for test cases generated at a 1.2 temperature consistently ranked high, with higher iteration values always at the top. 
However, the difference in estimator values converges after $i=5$.}



\resultbox{Higher temperature (1.2) and more prompt iterations contribute to healthier test cases, resulting in improved F-measure 96.6\% classifying correct and incorrect variants.}

%% file: RQ4.tex
\subsection{RQ5: Effectiveness of \toolit\label{sec:rq4}}
We conducted a comparative analysis to assess the effectiveness and novelty of \toolit in automating a wider range of code variations, particularly those that were previously challenging to achieve. 
For this evaluation, we compared \toolit with \textit{PyEvolve}, which is recognized as a leading tool for handling previously unseen variations, with a precision of 97\% and a recall of 94\% when automating transformations.
We applied \toolit to execute CPATs on new target sites and measured its capability to automate unseen variations belonging to both \VTone and \VTtwo. 
Simultaneously, we used \textit{PyEvolve} with the same input changes to determine its performance in automating unseen variations. 
This comparison allowed us to assess the impact and uniqueness of our novel contribution in automating a broader spectrum of code variations.

\subsubsection{\textbf{Dataset and Experimental Setup}}
To employ the \cpats, we utilize a selection of 200 top-tier projects that have been previously confirmed by researchers~\cite{rcaptminer2022ICSE} to showcase diversity in aspects such as developers, line counts, Python files, and star ratings. 
These projects include renowned Python libraries such as \textit{NLTK}, \textit{Keras}, and \textit{microsoft/DeepSpeed}. 
We configured \tool's parameters as discussed in Section \ref{sec:rq3_1} and \ref{sec:rq3_2} and applied the \cpats listed in Table \ref{table:cpat} to these projects. 
Our analysis focuses on quantifying the target codes that \tool automated but \textit{PyEvolve} could not, allowing us to comprehensively evaluate \tool's unique and advanced capabilities in automating a wider spectrum of code variations compared to \textit{PyEvolve}.

\subsubsection{\textbf{Results}}
\Cref{table:cpat} shows the applicable variations generated by \tool for each \cpat. 
It shows that \tool generates applicable variants, with an average ratio of \CPATtoVariatoin ranging from a minimum of 9 to a maximum of 110 per CPAT.
This empowers \tool to conduct an average of \CPATtoVariatoin more searches for applying the \cpat compared to \textit{PyEvolve}.
Building on this advantage, \Cref{table:cpat} shows the number of transformations carried out by \tool for each CPAT, along with those performed by \textit{PyEvolve}.
Particularly, \tool outperforms \textit{PyEvolve} by a minimum of \minOpportunitiesRatioOverBaseLine times across all CPATs, achieving a maximum ratio of \maxOpportunitiesRatioOverBaseLine times transformations compared to the baseline, resulting in an average increase of \opportunitiesRatioOverBaseLine in transformation opportunities. 
This demonstrates that \tool consistently outperforms \textit{PyEvolve} by a substantial margin across all evaluated CPATs. 
This superior performance is attributed to \tool's ability to generate and search for code variants more effectively, resulting in significantly higher opportunities for code improvement compared to the baseline.



\resultbox{
\tool is capable of generating applicable variations with an average ratio of 
\CPATtoVariatoin per CPAT, resulting in an average of \opportunitiesRatioOverBaseLine more transformations compared to the baseline.
}

%% file: RQ6.tex
\subsection{RQ6: Usefulness of \tool\label{sec:rq6}}
We observed that \tool demonstrates a remarkable ability to automate a broader range of code variations than the baseline \textit{PyEvolve}. 
However, it is important to examine the usefulness of the code transformations that \tool automates but \textit{PyEvolve} does not, especially for real-world developers.
To accomplish this, we submitted the patches generated solely by \tool to open-source projects and analyze the response of the developers.

\subsubsection{\textbf{Dataset and Experimental Setup}}
We selected \rqSixProjects high-quality projects, including prominent ones like \textit{Keras}, \textit{microsoft/DeepSpeed}, and \textit{NLTK} to apply the \cpats listed in \Cref{table:cpat}.
The inclusion of these professionally-maintained projects underscores \tool's ability to identify target codes, which only \tool can automate, potentially even those that expert programmers might overlook in their coding practices.
We executed both \tool and \textit{PyEvolve}, identifying transformations exclusively performed by \tool and then submitting them as pull requests to projects.

\subsubsection{\textbf{Results}}
\tool effectively transformed \pullCPATInstances instances of \cpats, leading to significant improvements in the performance and quality of the affected Python code. These patches updated \pullUpdatedFiles source code files and impacted \pullUpdatedLOC SLOC. 
Following the changes performed in each project, we  executed all the available test cases to ensure that the changes did not introduce regressions. 
Then, we notified the maintainers of the open-source projects through pull requests to incorporate our proposed changes.
Our pull requests received acknowledgment and approval even from maintainers of prestigious and extensively optimized codebases, such as those from Microsoft and IBM. 
In total, we submitted \numberOfPullReq pull requests, each containing \pullCPATInstances instances of \cpats. 
At the time of this writing, \acceptedInstances (with an acceptance ratio of \acceptedRatioOfCPATS) instances were accepted, while \rejectedPulls pull requests were declined. 
The remaining pull requests are still under review.
These positive responses demonstrate the practical usefulness of the program transformations made solely by \tool.


We discovered three major reasons for pull request rejections:
\begin{enumerate*}[label=(\roman*)]
\item NumPy, a popular library for numerical computing in Python, primarily runs on CPU and does not have native GPU support.
Because of that, a pull request submitted to \textit{dmlc/dgl}, which aimed to transform code into NumPy APIs that are optimized for GPUs, was rejected,

\item A patch submitted to \textit{HKUNLP/UnifiedSKG} was declined because they were hesitant to modify their methods in order to avoid disrupting their forks, and

\item The project \textit{netsharecmu/NetShare} opted not to accept changes related to a deprecated function that is scheduled for removal in the next release.
\end{enumerate*}

\tool demonstrated its capability by successfully optimizing code within selected projects that were already highly optimized and well-maintained. This achievement not only showcases \tool's considerable value but also emphasizes its ability to discover and leverage substantial opportunities for improvement in less frequently maintained projects. The widespread presence of these projects implies that \tool has a significant potential to make a meaningful impact on the broader development community. It offers far-reaching advantages by improving code quality, performance, and maintainability in a diverse range of Python applications.
\resultbox{
Developers accepted \acceptedRatioOfCPATS of the \pullCPATInstances instances, highlighting the usefulness of \tool's changes.
}

%% file: ThreatsToValidity.tex
\section{Threats To Validity}
\begin{enumerate}[wide, leftmargin=*,labelwidth=!, labelindent=0pt]
\item \textbf{Internal Validity}: \textit{Does our tool produce valid results?}
We ensure internal validity by relying on established tools like \textit{R-CPATMiner}, effective for mining \cpats, and \textit{PyEvolve}, which achieves 97\% precision and 94\% recall in code changes.

\item \textbf{External Validity}: \textit{Do our results generalize?} 
\toolit's effectiveness hinges on the chosen LLM model, which we address by evaluating variation generation among top LLMs and selecting the most suitable one. 
As LLMs upgrade, improved results can be anticipated. The design of \tool enables seamless adaptation to LLMs.
Moreover, the effectiveness of variation generation can vary depending on the selected \cpats, potentially limiting the generalizability of the analysis results to other \cpats. 
While this limitation could be mitigated by including more \cpats, the manual analysis conducted in evaluations hinders us from adding additional variations. 
To address this challenge, we selected \cpats that cover all four \cpat kinds discovered by Dilhara et al.~\cite{rcaptminer2022ICSE}.

{The techniques in \tool, designed for Python, are broadly applicable and conceptually robust across different programming languages. Their success in generating code variants largely hinges on the language's syntactic diversity and its library ecosystem's breadth, which provides similar functions via various APIs. Hence, while \tool's methods are universally applicable, their performance depends on the language's syntax and library API richness.}

\item \textbf{Verifiability}:
The data, source code, and executable of \tool are publicly available~\cite{artifacts}. 
\end{enumerate}

%% file: RelatedWork.tex
\section{Related Work}
We group the related work in two areas: 
\begin{enumerate*}[label=(\roman*)]
\item Inferring and applying changes using examples changes, and
\item LLMs for analysing and transforming source codes.
\end{enumerate*}

\textbf{LLMs for analysing and transforming source code:}
\textit{MELT}~\cite{ramos2023melt} is the closest related work that employs LLM to generate transformation examples from library pull requests, facilitating the adaptation of clients to new APIs. While their approach also utilizes LLM, like ours, their primary objective is to address the rule generalization issue~\cite{Xiang2021APIFix}, aiming for rules that are less specific. In contrast, our focus lies in generating \VTone and \VTtwo unseen variants. Although both approaches tackle two main, distinct challenges in TBE, we hypothesize that combining these two works could further enhance the number of transformations.
\textsc{CodeBERT}~\cite{feng-etal-2020-codebert}, CodeT5~\cite{wang-etal-2021-codet5}, \textsc{DeepCoder}~\cite{Balog2016DeepCoderLT}, \textsc{CodeX}~\cite{openai-codex},
\textsc{EM-Assist}~\cite{pomian2024together}
and \textsc{Synchromesh}~\cite{poesia2021synchromesh} are examples of LLM-based models that have shown promise in tasks such as code completion, code summarization, code refactoring and code generation. 
Additionally, LLMs have been used for code repair and bug detection, with tools like DeepBugs~\cite{deepbugs}, and MMAPR~\cite{zhang2022repairing} automatically identifying and fixing common programming errors.
Although these techniques have shown efficiency in various facets of source code transformations, our research concentrates on leveraging LLMs to broaden the applicability of TBE systems and transform new target code sites that are variants of the examples found in training. 
This significantly improves the code transformation potential of TBE systems.

\textbf{Inferring and applying changes using examples changes:}
Researchers have made significant improvements in the field of TBE~\cite{Meng2013Lase,rolim2017ReFazer,Andersen2010,inferrule,AppEvolveFazziniAST2019,pyevolve2022ICSE,Meditor:ICPC:2019,A3LamotheTSE2022,MengSystematicEditingPLDI2011}.
These techniques employ various approaches to increase the number of reliable code transformations per rule.
\textsc{Lase} \cite{Meng2013Lase}, \textsc{ReFazer} \cite{rolim2017ReFazer}, \textsc{Spdiff} \cite{Andersen2010}, \textsc{TCInfer} \cite{inferrule}, \textsc{AppEvolve} \cite{AppEvolveFazziniAST2019}, and \textsc{APIFix} \cite{Xiang2021APIFix} utilize multiple examples to generate transformation rules, identifying commonalities and differences among input examples to abstract adaptations.
The effectiveness of these rules increases with the number of input examples used to infer them.
In this regard, \textsc{APIFix}~\cite{Xiang2021APIFix} leverages existing codes to synthetically generate example code changes. 
These changes are then used as inputs to synthesis tools, enhancing the generalization of the rules. 
On the other hand, \textsc{PyEvolve} \cite{pyevolve2022ICSE} and \textsc{SpInfer} \cite{serrano2020spinfer} are the tools most closely related to \tool.
\textsc{SpInfer} \cite{serrano2020spinfer} has the ability to handle control flow variants by utilizing the "..." operator to represent random code within the input code. 
On the other hand, \textsc{PyEvolve} \cite{pyevolve2022ICSE} can accommodate data and control flow unseen variants. 
However, it is unable to handle the more complex \VTone or \VTtwo variants, which can be successfully transformed  by \tool.

{\textbf{Automating Python code changes:}
Researchers have studied Python idioms and their usage in Python systems. 
Phan-udom et al.~\cite{Phanudom2020Teddy} analyzed 58 non-idiomatic and 55 idiomatic changes, while Alexandru et al.~\cite{Alexandru2018idioms} provided a list of Python idioms from a developer survey.
Sakulniwat et al.~\cite{sakulniwat2019visualizing} studied the evolution of Python's \code{with} statements over time, while Wang et al.~\cite{PyNose2021ASEWang} examined Python code smells.
Recently, Zhang et al.~\cite{Zhang2022IdiomaticPython} employed AST rewriting to automatically refactor nine Python idioms. 
Despite these contributions, the literature tends to offer guidelines and insights rather than complete, automated solutions for such transformations, often limited by the constraints of predefined, static rules. 
By using \tool, we implement a technique to infer transformation rules and automate the process of identifying and implementing syntactically different yet semantically equivalent code transformations. 
\tool is poised to significantly benefit both Python and ML developers, who form a substantial portion of the Python community~\cite{braiek2018open,TechnicalDebtinML2021ICSE,software2}.}
